\documentclass{article}
\usepackage[preprint]{tmlr}        

\usepackage{amsmath, amssymb, amsthm}
\usepackage{booktabs}
\usepackage{algorithm}
\usepackage{algpseudocode}
\usepackage{microtype}
\usepackage{hyperref}
\usepackage[capitalize, noabbrev]{cleveref}
\usepackage{graphicx}
\usepackage{tikz}
\usetikzlibrary{arrows.meta, positioning, fit, shapes.geometric}
\usepackage{xcolor}

\newcommand{\rev}[1]{#1}

\newcommand{\bK}{\mathbf{K}}
\newcommand{\bV}{\mathbf{V}}
\newcommand{\bq}{\mathbf{q}}
\newcommand{\bk}{\mathbf{k}}
\newcommand{\R}{\mathbb{R}}
\newcommand{\rope}{R_{\theta}}
\newcommand{\BagMerge}{\textsc{BagMerge}}
\newcommand{\CanonicalMerge}{\textsc{CanonicalMerge}}
\newcommand{\LatentFragmentSet}{\textsc{LatentFragmentSet}}

\title{Cache Merging as a Convergent Replicated State for\\
       Multi-Agent Latent Reasoning}

\author{\name Carlos Baquero \\
        \addr Faculdade de Engenharia (FEUP), Universidade do Porto (UP), Portugal \\
        \texttt{cbm@fe.up.pt}
   \AND \name Lu\'is Brito \\
        \addr Escola Superior de Tecnologia e Gest\~ao (ESTG), Polit\'ecnico de Viana do Castelo (IPVC), Portugal \\
        \texttt{britoluis@estg.ipvc.pt}}

\begin{document}
\maketitle

\begin{abstract}
Multi-agent latent reasoning composes the KV-cache contributions of
multiple agents into a single context for a final agent. Recent work
(Agent Primitives) realises this composition by
concatenating per-agent caches along the sequence axis with RoPE
re-encoding, a construction we name \BagMerge{}. \BagMerge{} is
non-commutative, and which input ordering is best is not predictable
a priori: it shifts with the deployment regime, the latent-step
budget, and even the model scale. We make this cache exchange a
\emph{convergent replicated state}. First, \CanonicalMerge{} fixes the
layout: a content-determined ordering by mean K-norm at a middle
transformer layer renders the merged cache byte-identical under any
permutation of the inputs, verified algorithmically on synthetic
tensors (arbitrary arity $N \le 5$) and bit-for-bit on the real KV
state of Qwen3-1.7B ($28$ layers) and Qwen3-4B ($36$ layers). Second, we separate the replicated
\emph{state} from decode-time \emph{layout}: the durable object is a
set of content-addressed latent fragments whose merge is set union,
a state-based CvRDT (commutative, associative, idempotent, absorbing),
and \CanonicalMerge{} is its deterministic render. Because the
render is byte-equivalent, every $N = 2$ accuracy number is inherited
unchanged and re-delivered duplicate fragments are absorbed rather than
re-concatenated. On a partitioned-reasoning benchmark (Qwen3-1.7B),
\CanonicalMerge{} lands within $4$ percentage points of the best
\BagMerge{} ordering in every cell of a $12$-cell regime
$\times$ budget $\times$ ordering matrix and matches it without needing
to know which ordering is best, trading a small, statistically
insignificant accuracy margin for an unconditional structural guarantee;
a four-cell Qwen3-4B scale check preserves the same qualitative result\rev{, and on a second model family (Llama-3.1-8B) the ordering gap widens to $19$ points while \CanonicalMerge{} again tracks the best ordering}. \rev{On real multi-document QA (HotpotQA bridge-$k = 2$, $n = 200$; MuSiQue 2-hop) \CanonicalMerge{} is the best cache-level method without choosing an ordering, at a $7$-point F1 cost against shipping the text on HotpotQA and at parity on MuSiQue}, while the closest training-free output-fusion baseline
(PackLLM) loses by $45$ points at matched generation budget, placing
cache-level merging in a regime distinct from output-level fusion. Finally, at $k > 2$ we
delimit the approach: cache merge transports and colocates latent
traces but does not by itself compose them, motivating subsequent work.
\end{abstract}

\section{Introduction}
\label{sec:intro}

Multi-agent language-model systems increasingly reason in \emph{latent}
space: instead of exchanging natural-language messages, agents pass
intermediate KV-cache state directly, giving a downstream agent access to
richer, higher-bandwidth representations than decoded
text~\citep{zou2025latentmas,jin2026agentprimitives}, part of a rapidly
growing shift toward latent-space computation in language
models~\citep{yu2026latentspace}. This latent exchange
is realized by composing the per-agent caches into a single context that a
final agent attends over. As the number of contributing agents grows, and
as settings move toward parallel reasoning, recursive
delegation~\citep{yang2026recursivemas}, and federated inference, this
composition step becomes the load-bearing operation, yet the operator that
performs it has been left largely implicit.

The de facto standard for the parallel case, which we name \BagMerge{},
concatenates the agents' caches along the sequence axis and re-encodes their
positions with RoPE~\citep{jin2026agentprimitives}. Concatenation is
ordered, and the RoPE re-encoding makes that order matter: \BagMerge{} is
non-commutative, so which agent's fragment occupies the privileged
position-$0$ prefix changes the decoded answer. Worse, \emph{which} ordering
is best is not predictable in advance: we find it shifts with the
deployment regime, the latent-step budget, and even the model scale
(\cref{sec:experiments}), so a deployer cannot fix a safe ordering
without per-setting validation.

This is not merely a tuning nuisance; it signals a missing abstraction. In
the settings latent exchange is most attractive for (federated or
query-blind inference, where independent parties contribute pre-encoded
state without a coordinator and often without the query itself), the
composition must be order-independent, must converge to the same result
regardless of delivery schedule, and must tolerate the same fragment
arriving more than once (gossip, retry, re-broadcast). These are precisely
the guarantees that conflict-free replicated data types (CRDTs) provide for
distributed state~\citep{shapiro2011crdt,preguica2018crdts}, yet prior
latent-MAS cache composition offers none of them: it is order-sensitive,
schedule-dependent, and duplicates re-concatenate rather than absorb. The
problem we take up is how to make inter-agent cache exchange a
\emph{convergent replicated state}, well-defined, coordinator-free, and
robust to delivery, without sacrificing accuracy.

We address this in two steps. First, \CanonicalMerge{} fixes the layout by
\emph{content} rather than by caller order: it sorts the per-agent caches by
their mean key-norm at a single middle transformer layer, a
content-determined score that renders the merged cache byte-identical under
any permutation of the inputs (\cref{sec:method}). We verify this
commutativity algorithmically for up to five agents and bit-for-bit on the
real KV state of Qwen3-1.7B ($28$ layers) and Qwen3-4B ($36$ layers).
Second, we separate the replicated \emph{state} from its decode-time
\emph{layout}: the durable object is a set of content-addressed latent
fragments whose merge is set union, a state-based CvRDT that is
commutative, associative, idempotent, and absorbing by one-line proofs,
and \CanonicalMerge{} is its deterministic render (\cref{sec:method-state}).
Because the render is byte-equivalent to a canonical concatenation, every
accuracy result transfers unchanged, while re-delivered duplicate fragments
are absorbed rather than re-concatenated.

To isolate the merge operator empirically we introduce a
partitioned-reasoning benchmark whose problems split into
individually-insufficient, jointly-necessary fragments, so that a correct
answer \emph{requires} integrating both caches (\cref{sec:benchmark}).
Across a regime $\times$ budget $\times$ ordering matrix, \CanonicalMerge{}
matches the best \BagMerge{} ordering in every regime without needing to
know which ordering is best; a three-draw greedy replication confirms this
parity under paired-bootstrap, Holm-corrected tests, trading a small,
statistically insignificant accuracy margin for an unconditional structural
guarantee (\cref{sec:experiments}). \rev{On real multi-document QA (HotpotQA and MuSiQue)
\CanonicalMerge{} is the best cache-level method without choosing an
ordering, at a $7$-point F1 cost against shipping the fragments as text on
HotpotQA and at parity on MuSiQue}, while the closest
training-free output-fusion baseline, PackLLM, trails by $45$ points at
matched budget, placing cache-level merging in a regime distinct from
output-level fusion. A re-delivery experiment then makes idempotence
load-bearing: naive re-concatenation degrades monotonically and grows the
cache linearly in the number of deliveries, whereas the CvRDT render is
invariant. Finally, beyond two agents we delimit the approach, cache
merging transports and colocates latent traces but does not by itself
compose them, motivating subsequent work (\cref{sec:beyond-n2}).

\paragraph{Contributions.}
\begin{itemize}
\item We identify and characterize the \emph{directionality flaw} of
\BagMerge{}, the concat-with-RoPE cache operator implicit in prior
latent-MAS work: it is non-commutative, and the best ordering is
unpredictable across regime, latent budget, and model scale
(\cref{sec:method-bagmerge,sec:experiments}).
\item We introduce \CanonicalMerge{}, a content-determined cache layout
whose output is byte-identical under any permutation of the inputs,
verified algorithmically up to $N = 5$ and bit-for-bit on real Qwen3-1.7B
and 4B KV state (\cref{sec:method,app:byte-crdt-tests,app:realmodel-byte-equality}).
\item We reframe the merged cache as a \emph{convergent replicated state}: a
set-based CvRDT (\LatentFragmentSet{}) whose deterministic render is
\CanonicalMerge{}, giving commutativity, associativity, idempotence, and
duplicate-absorption on the state, with every $N = 2$ accuracy number
inherited byte-exactly (\cref{sec:method-state}).
\item We construct a partitioned-reasoning benchmark that isolates the merge
operator, and show under a multi-seed, Holm-corrected paired analysis that
\CanonicalMerge{} matches the best fixed ordering in each regime without
selecting it, and operates in a regime distinct from output-level fusion
(PackLLM $-45$~pp at matched budget) (\cref{sec:benchmark,sec:experiments}).
\item We demonstrate idempotence as an operational property under fragment
re-delivery, and delimit the approach at $k > 2$, merge transports but
does not compose, to motivate subsequent work
(\cref{sec:experiments-redeliver,sec:beyond-n2}).
\end{itemize}

\section{Related Work}
\label{sec:related}

\subsection{Latent reasoning and inter-agent latent communication}
\label{sec:related-lineage}
Latent chain-of-thought, introduced as Coconut~\citep{hao2024coconut},
replaces sampled text tokens with continuous last-layer hidden states fed
back autoregressively, decoupling internal computation from the discrete
vocabulary. LatentMAS~\citep{zou2025latentmas} extends this to the
multi-agent setting: rather than communicating in text, each agent's KV
cache is prepended to the next agent's working memory, forming a
\emph{sequential} chain of latent thoughts. Interlat~\citep{interlat2025}
shares the inter-agent latent-message hypothesis with a small adapter that
projects one agent's hidden states into the next's input space, and
RecursiveMAS~\citep{yang2026recursivemas} casts a heterogeneous system as a
recursive latent computation, transferring last-layer hidden states through
a learned residual link over recursion rounds. All of these compose agents
\emph{sequentially}, the next agent conditions on the union of prior
caches, and their composition is order-dependent by construction. We
study the \emph{parallel} variant: agents run independently on disjoint
fragments and a merge function combines their caches, a setting in which the
choice of merge function has consequences the sequential constructions never
expose. A recent taxonomy of latent inter-agent
communication~\citep{beyondtokens2026} catalogues the operators in use
(concatenation, prepending, arithmetic, cross-attention, cache restoration);
none provides the combination we study, a content-addressed set-union
state, a deterministic cache render, duplicate absorption, and byte-stable
input-order invariance, which is the cell our construction fills.

\subsection{Parallel reasoning and output-space fusion}
\label{sec:related-parallel}
Many techniques fan out to $N$ reasoning paths and aggregate in
\emph{answer space}. Self-Consistency~\citep{wang2023selfconsistency} samples
$N$ decodings and votes; Multi-Agent Debate~\citep{du2023debate} runs
role-conditioned agents through rounds of textual exchange; Tree of
Thoughts~\citep{yao2023tot} searches a tree with branch-and-evaluate
operators; Mixture of Agents~\citep{wang2024moa} cascades parallel proposers
and aggregators; Universal Self-Consistency~\citep{chen2023usc} generalises
voting to LLM-judged majority over free-form text. All discard the latent
state that produced each output and read text. \CanonicalMerge{} is the
cache-space analogue of the last: rather than asking a selector to pick the
most consistent generation, a choice with known position
bias~\citep{zheng2023judge}, we let the geometry of the K vectors
determine which thinker is primary and let the judger consume the merged
cache directly. Output-level \emph{fusion} aggregates per-thinker next-token
distributions instead; PackLLM~\citep{mavromatis2024packllm} is the closest
training-free neighbour we compare against (\cref{sec:experiments-packllm}),
and the newest training-free fusion methods~\citep{adafuse2026,unite2024}
remain strictly logit-level, never touching hidden state or KV, the
limitation on which cache-level merging separates from output fusion.

\subsection{\texorpdfstring{\BagMerge}{BagMerge}: concat-with-RoPE as prior art}
\label{sec:related-bagmerge}
Agent Primitives'~\citep{jin2026agentprimitives} ``Voting and Selection''
primitive runs $N$ parallel solvers and combines their KV caches by
concatenating along the sequence axis, re-rotating each successor's K vectors
so its positions extend the prefix; it gives the explicit formula
$R_\theta(t + n_B)\, R_\theta(t)^{-1}\bK_t$ and reports large gains at 8B/70B
over concatenation without re-encoding. It applies the merge at $N = 3$
solvers, but in a voting setting where each solver independently attempts the
\emph{full} task and a selector aggregates among complete candidates, so the
merge \emph{selects} rather than \emph{composes}; our partitioned benchmark
instead forces the merge to compose jointly-necessary fragments, which is where
the $k > 2$ ceiling appears (\cref{sec:beyond-n2}). LatentMAS~\citep{zou2025latentmas}
exchanges KV caches through the same prepend mechanism but in a \emph{fixed,
sequential} order (\cref{sec:related-lineage}), each agent conditioning on its
predecessor's cache, so it never confronts the parallel input-order choice; the
parallel, order-free construction is what we name \BagMerge{}, for its multiset
semantics. Its flaw, which Agent Primitives does not
measure, is that the prefix slot is asymmetric (its K vectors keep their
original rotation; later slots are re-rotated), so the choice of which cache
occupies it changes accuracy, a $6$-percentage-point gap between the two
orderings of $N = 2$ thinkers on our benchmark (\cref{sec:experiments}).
RoFormer~\citep{su2024roformer} anchors the rotation formalism and
YaRN~\citep{peng2024yarn} documents the degradation when K vectors appear at
out-of-distribution positions, explaining why the asymmetry is load-bearing.
Closest in architecture is
Parallel-Synthesis~\citep{liu2026parallelsynthesis}, in which a synthesizer
directly consumes the caches of $N$ parallel same-backbone workers via a
cache mapper that performs exactly this RoPE manoeuvre. It differs from
\CanonicalMerge{} on every point that constitutes our contribution: it
concatenates in \emph{fixed} worker order with no content-determined
ordering, commutativity analysis, or byte-identical-render claim, and it
\emph{requires post-training} (a cache mapper plus a synthesizer LoRA). The
authors report that their no-training variant ``performs substantially
worse'', an architectural neighbour whose need for learned calibration is
consistent with our $k > 2$ result (\cref{sec:beyond-n2}) that a free cache
merge transports but does not compose.

\subsection{Content-determined cache operators and CRDT structure}
\label{sec:related-crdt}
The CRDT literature~\citep{shapiro2011crdt,preguica2018crdts} distinguishes
operation-based (CmRDT) data types, whose operations commute, from
state-based (CvRDT) data types, which additionally form a semilattice with
idempotent, absorbing merge. The \emph{rendered} \CanonicalMerge{} operator
has the commutativity and byte-identity we need for one-shot rendering, but it
is not itself the replicated data type: it is not idempotent, because RoPE
re-rotation moves a re-delivered fragment's K vectors out of the positions
where the content function recognises them. We
recover the full property set by separating state from render
(\cref{sec:method-state}): the replicated state is \LatentFragmentSet{}, a
set of content-addressed unshifted fragments under set union, a
state-based CvRDT, of which \CanonicalMerge{} is the deterministic render.
The closest framing is Gillespie's CRDT for neural-network
merging~\citep{gillespie2026crdtmerge}, which independently pairs an OR-Set
state-CRDT with a canonical render; it merges \emph{persistent weights
offline}, however, with no positional structure, no per-inference state, and
no decode-time composition, whereas we merge \emph{per-inference latent KV}
fragments, to our knowledge the first CvRDT over inference-time latent
state. Gillespie pre-empts the broad claim ``first CRDT over neural state'';
our novelty is the inference-time KV object, the RoPE render, and the
empirical latent-merge boundary at $k > 2$. As a cache operator \CanonicalMerge{} is also kin to KV-cache
compression, StreamingLLM~\citep{xiao2024streamingllm},
H2O~\citep{zhang2023h2o}, and SnapKV~\citep{li2024snapkv} all let content
decide which K vectors are special, but \emph{within one cache}. A separate
line \emph{merges} rather than evicts redundant KV
entries~\citep{kvmerger2024}, but again within a single sequence's cache,
collapsing adjacent positions for memory; we merge whole caches \emph{across
agents} to combine content, an orthogonal axis.

\subsection{Federated / query-blind inference}
\label{sec:related-federated}
Federated \emph{training} is established~\citep{douillard2023diloco};
federated \emph{inference} is less explored. Retrieval-Augmented
Generation~\citep{lewis2020rag} is the canonical pattern of external sources
contributing fragments to a central generator, but in \emph{text} space;
\CanonicalMerge{} is its cache-space analogue when fragments are pre-encoded
as KV caches, and the encoding can be \emph{query-blind}, thinkers encode
without seeing the global query. The indirect-prompt-injection
literature~\citep{greshake2023prompt} notes that any pipeline where retrieved
data crosses the data/instruction boundary \emph{by position} rather than by
content is exploitable; \BagMerge{} has this property in cache space, and
\CanonicalMerge{}'s content-based primary selection gives a principled hook
it lacks.

\subsection{Implementation context}
\label{sec:related-impl}
PagedAttention and vLLM~\citep{kwon2023pagedattention} establish per-request
KV caches as first-class, block-addressable objects that can be shared,
copied, and recombined across requests; \CanonicalMerge{} slots into this
abstraction as another reduction over the same block-structured caches, with
no new systems primitive required.

\section{Partitioned-Reasoning Benchmark}
\label{sec:benchmark}

We construct a multi-step reasoning benchmark in which each problem
is \emph{partitioned} between two thinkers: each thinker receives one
of two textual fragments that, individually, contain insufficient
information to derive the answer but jointly determine a unique
integer answer. The merge operator under test is responsible for
combining the two per-thinker latent states into a coherent context
the judger can decode from.

\subsection{Motivation}
\label{sec:benchmark-motivation}

Existing multi-agent latent-reasoning
benchmarks~\citep{zou2025latentmas,jin2026agentprimitives} give each
thinker access to the global query. In such designs, any accuracy
lift from a merge operator is confounded with the extra inference
compute applied to the same problem. A same-data evaluation cannot
distinguish a merge that \emph{integrates} information across
thinkers from one that merely regularises the output of a single
thinker (full discussion in \cref{app:same-data-motivation}). The
partitioned construction eliminates the same-data confound by
design: a single thinker operating on its half can only achieve a
floor-level accuracy, because the missing fragment carries
information the model has no prior way to recover.

\subsection{Construction}
\label{sec:benchmark-construction}

The benchmark comprises $100$ problems drawn evenly from five problem
families, $20$ per family. Each problem has two text fragments
$t_a$ and $t_b$ and an integer answer. The fragments are constructed
so that any single fragment is insufficient.

\begin{table}[t]
\centering
\small
\begin{tabular}{lp{4.5cm}p{4.5cm}r}
\toprule
Family       & $t_a$                              & $t_b$                          & answer \\
\midrule
CONSTRAINT   & ``Alice is $3{\times}$ as old as Bob.''   & ``Together they are $40$.''      & $30$ \\
RECIPE       & ``One batch uses $2$ cups flour.''   & ``Make $5$ batches.''            & $10$ \\
DISCOUNT     & ``\$$10$ each, $20\%$ off for groups $\geq 4$.'' & ``$5$ friends buy tickets.''  & $40$ \\
NEONYM       & ``$1$ wog $= 7$ wibbles.''             & ``The race is $49$ wibbles.''    & $7$ \\
LOOKUP       & ``Red$=3$, Blue$=5$, Green$=2$ points.'' & ``$4$ red, $2$ blue, $6$ green.''    & $34$ \\
\bottomrule
\end{tabular}
\caption{Family schemas with one example shown per family; $20$
problems per family vary the surface numerical values while preserving
the schema.}
\label{tab:benchmark-families}
\end{table}

The NEONYM family uses fictional units (e.g.\ \texttt{wog},
\texttt{wibble}) for which the model has no prior knowledge; the
conversion factor must be obtained from $t_a$. The other four
families use surface forms a $1.7$B-parameter model could in
principle have memorised, but the specific numerical values vary
problem-by-problem.

We provide four data-loader views on each problem. The two-thinker
\texttt{split} view sets $\text{thinker}_0$'s input to $t_a$ and
$\text{thinker}_1$'s to $t_b$. Three single-thinker views serve as
controls: \texttt{full} concatenates $t_a + t_b$ as a ceiling, and
\texttt{a\_only} / \texttt{b\_only} expose just one fragment.

\subsection{Empirical validation}
\label{sec:benchmark-validation}

\begin{table}[t]
\centering
\small
\begin{tabular}{lrl}
\toprule
view              & accuracy            & interpretation \\
\midrule
\texttt{full} (C0)     & $100\%$        & ceiling --- single thinker on $t_a + t_b$ \\
\texttt{a\_only} (C1)  & $5\%$          & half $a$ alone is insufficient \\
\texttt{b\_only} (C2)  & $2\%$          & half $b$ alone is insufficient \\
\bottomrule
\end{tabular}
\caption{Single-thinker baselines on partitioned $k=2$
($100$ problems, Qwen3-1.7B). The headroom available for a
two-thinker merge to contribute is $95$~pp.}
\label{tab:benchmark-baselines}
\end{table}

The single-thinker baselines establish that the partitioning is
strong: an agent reading only $t_a$ scores $5\%$, an agent reading
only $t_b$ scores $2\%$, while a single agent reading $t_a + t_b$
scores $100\%$. The headroom available for a two-thinker merge to
contribute is therefore $\sim 95$ percentage points, and an
effective merge operator should land between the single-fragment
floor and the full-text ceiling. Numbers reported in
\cref{sec:experiments} are interpretable against this scale.

\subsection{Two regimes: query-known and query-blind}
\label{sec:benchmark-regimes}

We evaluate the merge operator under two deployment regimes that
correspond to qualitatively different real-world settings.

\paragraph{Query-known.} Each thinker receives both its data
fragment and the global query. This is the coordinated setting in
which a central component broadcasts the query to all thinkers
before they encode their fragments.

\paragraph{Query-blind (federated).} Each thinker receives only its
data fragment; the global query reaches only the judger, after the
merge. This is the federated setting in which each data source
encodes its fragment offline without coordinating on the query,
for instance, a multi-party inference setup where data shards
reside on hosts that do not see each other's queries.

The two regimes are distinguished by a single flag in our reference
implementation; the same $100$ partitioned-$k=2$ problems are used
for both. The two regimes produce qualitatively different K-norm
content signatures for the same thinker (\cref{fig:regimes}), and
the comparison between them is the central piece of evidence in
\cref{sec:experiments} for the \emph{structural robustness} of the
merge operator we propose.

\begin{figure}[t]
\centering
\begin{tikzpicture}[
  font=\small,
  frag/.style={align=center, font=\footnotesize\itshape},
  box/.style={draw, rounded corners, minimum width=1.5cm,
              minimum height=0.62cm, align=center},
  merge/.style={draw, rounded corners, fill=blue!8, minimum width=1.25cm,
                minimum height=0.62cm, align=center},
  jud/.style={draw, rounded corners, fill=blue!8, minimum width=1.4cm,
              minimum height=0.62cm, align=center},
  qnode/.style={draw, ellipse, fill=gray!12, inner sep=2.5pt, align=center},
  flow/.style={-{Latex}, thick},
  qflow/.style={-{Latex}, dashed, gray!55!black},
]

\begin{scope}[local bounding box=A]
  \node[frag] (taA) at (-0.3,0.62)  {$t_a$};
  \node[frag] (tbA) at (-0.3,-0.62) {$t_b$};
  \node[box]  (t0A) at (1.05,0.62)  {Thinker 0};
  \node[box]  (t1A) at (1.05,-0.62) {Thinker 1};
  \node[merge](mA)  at (3.0,0)       {Merge};
  \node[jud]  (jA)  at (4.7,0)       {Judger};
  \node       (ansA) at (6.3,0)      {answer};
  \node[qnode] (qA) at (3.0,2.0)     {query $Q$};

  \draw[flow] (taA) -- (t0A);
  \draw[flow] (tbA) -- (t1A);
  \draw[flow] (t0A.east) -- (mA);
  \draw[flow] (t1A.east) -- (mA);
  \draw[flow] (mA) -- (jA);
  \draw[flow] (jA) -- (ansA);
  \draw[qflow] (qA) to[out=165,in=80] (t0A.north);
  \draw[qflow] (qA) .. controls (0.7,2.7) and (-0.9,2) .. (t1A.north west);
  \draw[qflow] (qA) to[out=15,in=90]  (jA.north);
\end{scope}

\begin{scope}[local bounding box=B, yshift=-3.4cm]
  \node[frag] (taB) at (-0.3,0.62)  {$t_a$};
  \node[frag] (tbB) at (-0.3,-0.62) {$t_b$};
  \node[box]  (t0B) at (1.05,0.62)  {Thinker 0};
  \node[box]  (t1B) at (1.05,-0.62) {Thinker 1};
  \node[merge](mB)  at (3.0,0)       {Merge};
  \node[jud]  (jB)  at (4.7,0)       {Judger};
  \node       (ansB) at (6.3,0)      {answer};
  \node[qnode] (qB) at (3.0,1.75)    {query $Q$};

  \draw[flow] (taB) -- (t0B);
  \draw[flow] (tbB) -- (t1B);
  \draw[flow] (t0B.east) -- (mB);
  \draw[flow] (t1B.east) -- (mB);
  \draw[flow] (mB) -- (jB);
  \draw[flow] (jB) -- (ansB);
  \draw[qflow] (qB) to[out=0,in=90] (jB.north);
\end{scope}

\node[anchor=base west, font=\small\bfseries] at (-1.7,0)    {(a)};
\node[anchor=base west, font=\small]          at (-1.7,-0.3) {query-};
\node[anchor=base west, font=\small]          at (-1.7,-0.6) {known};
\node[anchor=base west, font=\small\bfseries] at (-1.7,-3.4) {(b)};
\node[anchor=base west, font=\small]          at (-1.7,-3.7) {query-};
\node[anchor=base west, font=\small]          at (-1.7,-4.0) {blind};
\end{tikzpicture}
\caption{The two evaluation regimes. The KV-cache flow (solid
arrows), thinkers $\to$ \textsc{Merge} $\to$ \textsc{Judger}
$\to$ answer, is identical across regimes; only the routing of
the global query $Q$ (dashed arrows) differs. In \textbf{(a)
query-known}, $Q$ is broadcast to both thinkers and the judger, so
each thinker encodes its fragment $t_i$ conditioned on the query.
In \textbf{(b) query-blind (federated)}, $Q$ reaches only the
judger, and each thinker encodes its fragment offline without
seeing the query. The same merge operator and the same $100$
partitioned-$k=2$ problems are used in both regimes; the
difference shifts the K-norm content signature each thinker
produces (\cref{sec:experiments-mechanism}).}
\label{fig:regimes}
\end{figure}

\section{Method}
\label{sec:method}

\subsection{Setup and notation}
\label{sec:method-setup}

Let $M$ be a decoder-only transformer with $L$ layers, $H$ attention
heads per layer, head dimension $d$, and rotary position embedding
(RoPE)~\citep{su2024roformer} with base $\theta$. For an input
sequence of $T$ tokens, $M$ maintains a per-layer KV cache
\[
\mathcal{C}
\;=\;
\bigl\{ (\bK^{(\ell)},\, \bV^{(\ell)}) \bigr\}_{\ell=1}^{L},
\qquad
\bK^{(\ell)}, \bV^{(\ell)} \in \R^{1 \times H \times T \times d}.
\]
The K vectors are stored \emph{post-rotation}: position $t$'s key has
been multiplied by the rotation matrix $\rope(t)$ defined elementwise
on the head dimension. Attention computes
$\bq_s^\top \bk_t = \bq_s^\top \rope(t - s) \bk_t$ under the RoPE
construction, so relative displacement is encoded automatically.

We consider $N$ parallel \emph{thinker} agents, each running $M$
independently on a fragment of the problem (\cref{sec:benchmark}) for
the same number of \emph{latent autoregressive
steps}~\citep{hao2024coconut}. Our partitioned construction places exactly
one fragment with each thinker, so the number of thinkers $N$ equals the
number of fragments $k$; we use $N$ and $k$ interchangeably, writing the
$N = 2$ case as ``$k = 2$'' when emphasising the partition.
Thinker $i \in \{1, \dots, N\}$ produces a KV cache
$\mathcal{C}_i = \{(\bK_i^{(\ell)}, \bV_i^{(\ell)})\}_\ell$
of length $T_i$. A \emph{judger} agent then consumes a merged cache
$\mathcal{C}^\star = \operatorname{Merge}(\mathcal{C}_1, \dots, \mathcal{C}_N)$
and decodes the answer with the same model $M$. The two roles are
asymmetric: the thinkers run latent steps and emit no text,
contributing only their KV caches, whereas the judger runs no latent
steps of its own and instead decodes the answer autoregressively with
$\mathcal{C}^\star$ as its KV prefix. No model parameters
are trained; the construction is training-free. The merge function
is the focus of this paper.

\subsection{\texorpdfstring{\BagMerge}{BagMerge} (prior work)}
\label{sec:method-bagmerge}

Agent Primitives~\citep{jin2026agentprimitives} defines a merge
operator that concatenates per-thinker caches along the sequence axis,
with RoPE re-encoding so successor positions extend the prefix range
without overlap. (LatentMAS~\citep{zou2025latentmas} exchanges caches
through the same prepend mechanism, but in a fixed sequential order,
each agent conditioning on its predecessor's cache rather than choosing a
concatenation order among parallel peers; \cref{sec:related-bagmerge}.)
We name this parallel construction \BagMerge{} (multiset semantics,
input-order preserving).

Let $\pi$ be a permutation of $\{1, \dots, N\}$ specifying the
concatenation order, and define cumulative offsets
$s_{\pi(1)} = 0$ and
$s_{\pi(i)} = s_{\pi(i-1)} + T_{\pi(i-1)}$ for $i > 1$. For each
layer $\ell$,
\begin{align}
\bK^{\star,(\ell)}_{\BagMerge}
&\;=\;
\operatorname*{cat}_{i=1}^{N}
\Bigl[\, \rope\!\bigl(s_{\pi(i)}\bigr) \rope(0)^{-1}\, \bK^{(\ell)}_{\pi(i)} \Bigr],
\label{eq:bagmerge-k}
\\
\bV^{\star,(\ell)}_{\BagMerge}
&\;=\;
\operatorname*{cat}_{i=1}^{N}
\Bigl[\, \bV^{(\ell)}_{\pi(i)} \Bigr].
\label{eq:bagmerge-v}
\end{align}
The values $\bV$ are not rotated and pass through unchanged. The
factor $\rope(s_{\pi(i)}) \rope(0)^{-1}$ shifts thinker $\pi(i)$'s
effective positions from $\{0, \dots, T_{\pi(i)}-1\}$ to
$\{s_{\pi(i)}, \dots, s_{\pi(i)} + T_{\pi(i)}-1\}$,
the formula given in \citet{jin2026agentprimitives}~\S 3.1 as
``KV PoPE''.

\paragraph{The directionality flaw.}
The judger's decoded outputs depend on $\pi$, i.e.\ the chosen order
of concatenation among the thinker outputs. This introduces an asymmetry, since the position-$0$
prefix slot is special: its K vectors keep their original RoPE
rotation, matching the geometry the judger's parameters were trained
on. Tail slots have been re-rotated by an amount that depends on the
prefix length $s_{\pi(i)}$. Empirically, this asymmetry causes a
$6$-percentage-point accuracy gap on our partitioned benchmark between
the two orderings of $N = 2$ thinkers (\cref{sec:experiments}).
Neither prior work measures this gap, nor proposes a principled
mechanism for choosing $\pi$.

\rev{RoPE itself is not flawed, and in ordinary single-context use there
is no choice to make, since token order is determined by the input
sequence. The sensitivity arises when independently encoded caches are
merged: the operator must assign one fragment to the position-$0$ prefix
range and the rest to shifted tail ranges, and that assignment is an
arbitrary deployment choice with no ground truth. The choice matters for
two reasons. Geometrically, tail keys are mapped to positions they never
occupied during encoding and are read across a longer relative
displacement, where decoder LMs show systematic position biases:
attention concentrates on early ``sink''
positions~\citep{xiao2024attentionsinks} and mid-context content is used
less reliably~\citep{liu2023lostmiddle}. Semantically, the fragments are
parallel contributions with no correct order, and \cref{sec:experiments}
shows the empirically better order is not predictable a priori; it
inverts between regimes and shifts with budget and scale, so no fixed
convention removes the sensitivity. This motivates a content-determined
order. \Cref{fig:attn-mass} measures the slot asymmetry directly: the
same fragment bytes, merged into the prefix slot versus the tail slot,
receive systematically different attention from the judger at every
layer.}

\rev{\begin{figure}[t]
\centering
\includegraphics[width=0.72\linewidth]{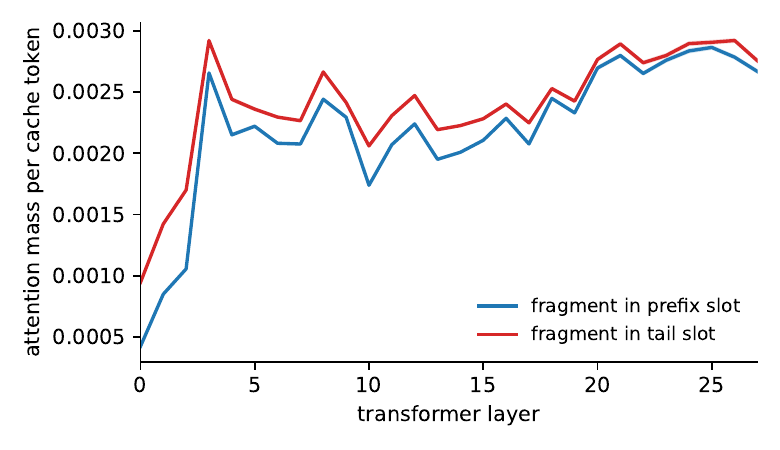}
\caption{Slot asymmetry measured at the mechanism level. For a fixed pair
of thinker caches, both \BagMerge{} orders are rendered and the judger's
prompt is forwarded over each merged cache; the curves show the per-token
attention mass on a fragment block when the same fragment occupies the
prefix slot (blue) versus the tail slot (red). Mean over $10$ problems
(two per family) and both fragments, $\pm 1$ standard error (bands are
narrower than the line width); Qwen3-1.7B, query-blind, $40$ latent
steps. The same bytes receive $8.5\%$ less per-token attention in the
prefix slot (pooled ratio $0.915$; the final query position alone gives
$0.920$). The tail block is closer to the judger's positions and draws
more raw attention, while the prefix block keeps its original rotation
geometry (\cref{sec:method-bagmerge}); the net effect on accuracy is
regime-dependent (\cref{sec:experiments}), which is why no fixed order
wins consistently.}
\label{fig:attn-mass}
\end{figure}}

\subsection{\CanonicalMerge}
\label{sec:method-canonical}

We replace input-determined permutation with a \emph{content-determined}
permutation derived from the K vectors themselves.

\paragraph{Definition.}
Fix a \emph{routing layer} $\ell^\star \in \{1, \dots, L\}$ (we use
$\ell^\star = 15$ for Qwen3-1.7B; \cref{sec:experiments} provides a
sensitivity sweep). For each thinker $i$, compute the mean K-norm over
the last $T^\star = \min_j T_j$ positions (tail-aligned to the
shortest cache):
\begin{equation}
\rho_i \;=\;
\frac{1}{T^\star}
\sum_{t = T_i - T^\star + 1}^{T_i}
\bigl\| \bK_i^{(\ell^\star)}[\,\cdot,\,\cdot,\,t,\,\cdot\,] \bigr\|_2
\quad
\text{(averaged over batch and heads).}
\label{eq:rho}
\end{equation}
Sort the thinkers in descending order of $\rho$, breaking ties by a
content hash $h_i$ over fragment $i$'s \emph{full} content (the shapes,
dtypes, and raw bytes of every layer's $\bK$ and $\bV$), identical to the
fragment id used by \LatentFragmentSet{} (\cref{sec:method-state}), so the
sort is a total deterministic order over distinct fragments, modulo the usual
cryptographic collision assumption.
The resulting permutation $\pi^\star$ is the \emph{canonical order}.
Apply \BagMerge{} with $\pi = \pi^\star$.
\Cref{alg:canonicalmerge} gives the procedure.
\rev{\CanonicalMerge{} fixes the layout deterministically, but the
rendered operator itself is \emph{not} idempotent; idempotence and
absorption are properties of the fragment-set state introduced in
\cref{sec:method-state}.}

\begin{algorithm}[t]
\caption{\CanonicalMerge}
\label{alg:canonicalmerge}
\begin{algorithmic}[1]
\Require Caches $\mathcal{C}_1, \dots, \mathcal{C}_N$; routing layer $\ell^\star$.
\For{$i = 1, \dots, N$}
  \State $\rho_i \gets \text{mean}\bigl(\bigl\| \bK_i^{(\ell^\star)}[\cdot, \cdot, \text{last } T^\star, :] \bigr\|_2\bigr)$
  \State $h_i \gets \operatorname{hash}\bigl(\text{full content of } \mathcal{C}_i\bigr)$ \Comment{all layers' $\bK, \bV$; the \LatentFragmentSet{} id}
\EndFor
\State $\pi^\star \gets \operatorname{argsort}_i\, \bigl(-\rho_i,\, -h_i\bigr)$ \Comment{descending; tiebreak by hash}
\State \Return $\BagMerge(\mathcal{C}_{\pi^\star(1)}, \dots, \mathcal{C}_{\pi^\star(N)})$
\end{algorithmic}
\end{algorithm}

\CanonicalMerge{} has \textbf{a single design parameter}, the routing layer
$\ell^\star$, one fixed architectural choice, which we report together with
a sensitivity sweep (\cref{sec:experiments-layer}) and re-select by a coarse
dev-set sweep when changing model scale (\cref{sec:experiments-scale}); the
full-content hash tiebreak is deterministic and parameter-free. There is no
fusion weight, similarity threshold, or per-task tuning.

The position-$0$ prefix slot is the privileged one: its K vectors retain
their original RoPE rotation and are read by the judger in the geometry its
parameters were trained on, whereas the tail thinker's vectors are
re-rotated to a shifted position range
(\cref{sec:method-bagmerge}). Sorting in \emph{descending} $\rho$ therefore
places the more informationally dense fragment, the one more likely to
carry the binding constraint, in the slot the judger reads most reliably,
and leaves the less critical fragment to absorb the slight geometric
degradation of the re-rotated tail. Where \BagMerge{} assigns this privileged
slot by arbitrary input order, \CanonicalMerge{} assigns it by content, per
problem.

\paragraph{Two properties from one mechanism.}

\emph{Commutativity at $N=2$.} The score function $\rho_i$ depends
only on thinker $i$'s K vectors, not on the input order. The sort is
total and content-determined; with a deterministic tiebreaker, the
permutation $\pi^\star$ is invariant under any reordering of the
input list. Therefore
\begin{equation}
\CanonicalMerge(\mathcal{C}_{\sigma(1)}, \dots, \mathcal{C}_{\sigma(N)})
\;=\;
\CanonicalMerge(\mathcal{C}_1, \dots, \mathcal{C}_N)
\label{eq:canonical-commute}
\end{equation}
bit-for-bit, for every permutation $\sigma$. The merged cache is
\emph{byte-identical} under thinker reordering, a structural
property verified algorithmically on synthetic tensors at
$N \in \{2, 3, 4, 5\}$ across $152$ permutation pairs
(\cref{app:byte-crdt-tests}) and bit-for-bit on the actual model's
$28$-layer KV state under the headline configuration
(\cref{app:realmodel-byte-equality}). The property is established by
construction; it does not depend on sample size or empirical
evaluation. We claim commutativity and byte-identity for the $N=2$
case studied in this paper. The empirical question of whether
the model integrates merged caches at $N > 2$ is deferred to
\cref{sec:beyond-n2}.

\emph{Per-problem adaptive routing.} The mean K-norm at a mid-network
layer correlates with the informational density of the fragment a
thinker has encoded, a claim grounded in the interpretability
literature on decoder-only
LMs~\citep{tenney2019,geva2021,belrose2023tunedlens}. For some
problems the binding constraint is in one fragment; for others, in
another. \CanonicalMerge{} selects the higher-density thinker for the
prefix slot \emph{per problem} rather than globally, subsuming the
$\pi = \text{id}$ and $\pi = \text{swap}$ orderings of \BagMerge{}
into a single content-routed choice. As we show in
\cref{sec:experiments}, this content function works in the
query-known regime, where each thinker has done question-conditional
reasoning; in the query-blind / federated regime the K-norm signal
flattens, but the commutativity property holds unchanged.

\subsection{From rendered operator to replicated state}
\label{sec:method-state}

\CanonicalMerge{} fixes the directionality of the \emph{rendered} cache, but
the rendered operator is not by itself a convergent replicated data type.
Re-merging a fragment already present in an intermediate result changes that
result: the copy of $B$ inside $\CanonicalMerge(A, B)$ has had its K vectors
RoPE-shifted out of their original positions, so the content-determined score
no longer recognises them as $B$, and a second merge does not absorb,
$\CanonicalMerge(\CanonicalMerge(A, B), B) \neq \CanonicalMerge(A, B)$. The
obstruction is that layout has been baked into the state.

We remove it by separating the replicated \emph{state} from the decode-time
\emph{layout}. The state is a set $\mathcal{S} = \{f_1, \dots, f_m\}$ of
\emph{content-addressed latent fragments}: each fragment is one thinker's
\emph{unshifted} legacy KV cache, and its identity is the full canonical byte
string of all layers' $\bK$ and $\bV$ (the hash is an address for that
identity, assumed collision-resistant; two distinct byte strings are distinct
fragments). Merge is set union,
\begin{equation}
\mathcal{S}_1 \sqcup \mathcal{S}_2 \;=\; \mathcal{S}_1 \cup \mathcal{S}_2,
\label{eq:setunion}
\end{equation}
which is commutative, associative, idempotent, and absorbing by the
corresponding one-line properties of $\cup$. Order the states by inclusion
over fragment identities, $\mathcal{S} \preceq \mathcal{T}$ iff every
identity in $\mathcal{S}$ is in $\mathcal{T}$; a local update only adds a
fragment, so it is inflationary ($\mathcal{S} \preceq \mathcal{S} \sqcup
\mathcal{S}'$), and union is the least upper bound under $\preceq$. The state
therefore forms a join-semilattice and is a state-based
CvRDT~\citep{shapiro2011crdt}; we call this object \LatentFragmentSet{}.

A \emph{render} is a deterministic pure function from a set to a single
decodable cache. We take render to be \CanonicalMerge{} applied to the
fragments in content-canonical order; because \CanonicalMerge{} is itself
permutation-invariant (\cref{eq:canonical-commute}), the render is a
well-defined function of the \emph{set}, not of any enumeration of it.
Crucially, $\mathrm{render}(\{A, B\})$ is byte-identical to the
\CanonicalMerge{} output on the same two caches, so every $N = 2$ result in
\cref{sec:experiments} is inherited unchanged: the state/render split adds the
CvRDT guarantees without altering a single decoded byte.

\emph{The algebraic guarantees live on the state, not on the operator.}
Because RoPE layout is applied only at render time, no merge ever has to
re-recognise content whose K vectors have already been rotated to a tail
slot; the absorption and idempotence that the rendered operator lacks hold
exactly for the set. We verify this behaviourally on real Qwen3 caches: under
re-delivery of a byte-identical fragment, \LatentFragmentSet{} absorbs the
duplicate and the rendered cache is invariant, whereas naive re-rendering of
the duplicated multiset grows the cache and can change the decoded answer
(\cref{sec:experiments-redeliver}).

One boundary is worth stating plainly. The absorption is \emph{syntactic}:
byte-identical copies hash equal and collapse to one element, but two caches
that encode the same information through different bytes are distinct elements
of the set. A semantically convergent join, one that also merges
near-duplicate but non-identical fragments, is left to future work.

\rev{\paragraph{Deployment preconditions.} The exact byte-level guarantees
above hold under conditions a federated deployment must provide: replicas
exchange \emph{byte-identical} fragments produced by matched model weights
under deterministic execution, and render with matched numerical kernels
(the IEEE-754 caveat of \cref{sec:limitations}). In heterogeneous
deployments, syntactically identical semantic content need not hash
identically, and the same fragment re-encoded on different hardware or
kernel versions may differ in low-order bits; such copies are distinct set
elements and are not absorbed. Convergence of the \emph{state} (set union
over whatever fragments exist) is unaffected; what weakens is deduplication
across replicas that do not share bytes.}

\subsection{Implementation}
\label{sec:method-impl}

\Cref{alg:latentfragmentset} collects the three operations of the
state/render split in one place. A fragment's identity \textsc{Id} is the
hash of its \emph{unshifted} per-layer $\bK, \bV$ bytes; local delivery
(\textsc{Ingest}) and cross-replica \textsc{Merge} are both set union keyed by
that identity, so re-delivered duplicates absorb (\cref{eq:setunion}); and
\textsc{Render} applies \CanonicalMerge{} (\cref{alg:canonicalmerge}) to the
fragments of the set, well-defined because \CanonicalMerge{} is
permutation-invariant. The hash and the merge act on state; the RoPE layout is
introduced only inside \textsc{Render}, which is what keeps idempotence and
absorption on the set rather than on the operator.

\begin{algorithm}[t]
\caption{\LatentFragmentSet{}: state operations and render}
\label{alg:latentfragmentset}
\begin{algorithmic}[1]
\Function{Id}{$\mathcal{C}$} \Comment{content address of a fragment}
  \State \Return $\operatorname{hash}\bigl(\operatorname{bytes}(\bK^{(1)}, \bV^{(1)}, \dots, \bK^{(L)}, \bV^{(L)})\bigr)$ \Comment{unshifted; collision-resistant}
\EndFunction
\Statex
\Function{Ingest}{$\mathcal{S}, \mathcal{C}$} \Comment{deliver a cache into the state}
  \State \Return $\mathcal{S} \cup \bigl\{(\Call{Id}{\mathcal{C}},\, \mathcal{C})\bigr\}$ \Comment{keyed by \textsc{Id}; a duplicate is absorbed}
\EndFunction
\Statex
\Function{Merge}{$\mathcal{S}_1, \mathcal{S}_2$} \Comment{CvRDT join $\sqcup$}
  \State \Return $\mathcal{S}_1 \cup \mathcal{S}_2$ \Comment{commutative, associative, idempotent, absorbing}
\EndFunction
\Statex
\Function{Render}{$\mathcal{S}$} \Comment{deterministic decode-time layout}
  \State \Return $\CanonicalMerge\bigl(\{\mathcal{C} : (\cdot, \mathcal{C}) \in \mathcal{S}\}\bigr)$ \Comment{Alg.~\ref{alg:canonicalmerge}; pure function of the set}
\EndFunction
\end{algorithmic}
\end{algorithm}

Reference implementation: strategy \texttt{novelty\_concat\_sym\_layout}
for $N = 2$ and \texttt{novelty\_concat\_sym\_layout\_n} for arbitrary
$N$. The merge operates on the standard HuggingFace legacy KV cache
format (per-layer tuples of $(\bK, \bV)$ tensors). The construction is
compatible with PagedAttention-style block-addressable
caches~\citep{kwon2023pagedattention}: the merge is a permutation over
already-allocated blocks plus a position-dependent RoPE-rotation
kernel that is $\mathcal{O}(T \cdot d)$ per thinker per layer,
negligible compared to the generation cost of the latent reasoning
steps that produced the cache.

\rev{\paragraph{Measured costs.} On the headline configuration
(Qwen3-1.7B, $40$ latent steps; medians over five problems, Apple M3 /
\textsc{mps}; \texttt{scripts/measure\_costs.py}): a fragment of $141$
cache tokens is a $16.2$~MB payload (all layers' $\bK, \bV$ in bfloat16),
its SHA-1 content hash costs $51$~ms, and the full \CanonicalMerge{}
render (K-norm scoring, canonical-order concatenation, tail re-rotation)
costs $142$~ms; a single $256$-token judger decode takes $16.4$~s, so
hash plus render amount to $1.2\%$ of one decode. The honest cost of
latent exchange is bandwidth, not compute: the same fragment as text is
under a kilobyte, so shipping raw KV is roughly four orders of magnitude
larger. Payload compression is orthogonal to the convergence construction
(the state is keyed by fragment identity) and is left to future work.}

\section{Experiments}
\label{sec:experiments}

Unless noted otherwise, experiments use Qwen3-1.7B with the routing
layer fixed at $\ell^\star = 15$ (a layer-sensitivity sweep in
\cref{sec:experiments-layer} justifies the choice);
\cref{sec:experiments-scale} replicates the headline
comparison on Qwen3-4B. Unless noted otherwise, each cell is $100$
problems from the partitioned-$k=2$ benchmark of \cref{sec:benchmark}
(\cref{sec:experiments-hotpotqa} uses $50$ HotpotQA problems and the
4B layer sweep uses $50$), sampling temperature $0.7$ with
$\texttt{top\_p} = 0.95$ and a single seed; the $N = 2$ accuracy matrix
is additionally replicated under greedy decoding across three problem
draws (\cref{app:bootstrap}).
Confidence intervals and paired tests are reported in
\cref{sec:experiments-stats} with full per-comparison tables in
\cref{app:bootstrap}.

\subsection{Headline: accuracy across regimes, budgets, and orderings}
\label{sec:experiments-headline}

\Cref{tab:headline-matrix} reports accuracy across the full 2~regimes
$\times$ 2~latent-step budgets $\times$ 3~methods matrix: query-known
vs.\ query-blind regime (\cref{sec:benchmark-regimes}), $20$ vs.\ $40$
latent reasoning steps per thinker, and BagMerge with $\pi = \text{id}$
(``default'') vs.\ BagMerge with $\pi = \text{swap}$ vs.\
\CanonicalMerge{}.

\begin{table}[t]
\centering
\small
\begin{tabular}{lcccc}
\toprule
Method                          & q-known @ $20$ & q-known @ $40$ & q-blind @ $20$ & q-blind @ $40$ \\
\midrule
\BagMerge{} default             & $0.89$         & $0.85$         & $\mathbf{0.90}$ & $\mathbf{0.85}$ \\
\BagMerge{} swap                & $\mathbf{0.90}$ & $\mathbf{0.91}$ & $0.83$        & $0.77$         \\
\emph{best \BagMerge}           & \emph{swap}    & \emph{swap}    & \emph{default} & \emph{default} \\
\midrule
\textbf{\CanonicalMerge{}}      & $\mathbf{0.93}$ & $\mathbf{0.94}$ & $0.86$         & $0.85$         \\
$\Delta$ (CM $-$ best \BagMerge) & $+0.03$       & $+0.03$        & $-0.04$        & $0.00$         \\
\bottomrule
\end{tabular}
\caption{Headline matrix on partitioned $k = 2$, $100$ problems each.
Best \BagMerge{} ordering per column in bold. The \emph{best
\BagMerge} row records which fixed $\pi$ achieves it; the $\Delta$
row records the \CanonicalMerge{} accuracy minus the best \BagMerge{}
accuracy in the same column.}
\label{tab:headline-matrix}
\end{table}

Two structural observations dominate the matrix.

\textbf{(i) The best \BagMerge{} ordering is regime-and-budget-conditional.}
In the query-known regime \BagMerge{} swap is best at both latent
budgets ($0.90$ at $20$ steps; $0.91$ at $40$); in the query-blind
regime \BagMerge{} default is best at both budgets ($0.90$ at $20$;
$0.85$ at $40$). The optimal ordering inverts between the two regimes
and the swap/default gap widens with budget: a deployer using
\BagMerge{} would need to know both the regime (does the data source
see the query?) and the latent budget to pick the right $\pi$, and
$7$--$14$ percentage points are at stake in the worst cells. The
dependence is in fact worse than ``regime $\times$ budget'': it is also
\emph{model}-dependent: on Qwen3-4B the inversion disappears and
swap wins both regimes (\cref{sec:experiments-scale}), so the best ordering
is not stable across the observed regimes and model scales; a deployer
cannot pick it without validation or prior regime knowledge. Agent
Primitives~\citep{jin2026agentprimitives}, which introduces this parallel
merge, does not measure the dependence, and a coordinator that did want to
specify $\pi$ correctly would need an oracle for it.

\textbf{(ii) \CanonicalMerge{} removes the ordering choice, and stays close
in the other cells.} The output is byte-identical under any permutation of the
inputs \emph{by construction} (\cref{sec:method-canonical},
\cref{eq:canonical-commute}), and the prefix occupant is content-routed per
problem rather than globally fixed. The remaining two knobs, regime and
latent budget, change the per-thinker caches themselves, so no merge can be
byte-invariant to them; what we observe there is instead \emph{empirical}
robustness, that \CanonicalMerge{}'s quality stays close to the best fixed
ordering. It lands within $4$ percentage points of the best \BagMerge{}
ordering in every cell and is statistically indistinguishable from it
($\Delta = +0.03 / +0.03 / 0.00 / -0.04$; every gap within the bootstrap CIs,
\cref{sec:experiments-stats}), matching the best ordering without needing to
know which ordering is best. The one cell in which \CanonicalMerge{}
underperforms is the query-blind, $20$-step cell: there \BagMerge{} default
happens to align with the content-density signal at $0.90$, but a deployer
would not know in advance which \BagMerge{} ordering will land there without
running both and comparing.

\paragraph{What this paper claims, narrowly.} We separate two kinds of
robustness and claim them differently. \emph{Structural / algebraic (exact, by
construction):} \CanonicalMerge{} is bit-for-bit invariant to input ordering
(\cref{sec:method-canonical}, with algorithmic verification in
\cref{app:byte-crdt-tests} and real-model verification on the Qwen3-1.7B KV
state in \cref{app:realmodel-byte-equality}), and the \LatentFragmentSet{}
state absorbs duplicate re-deliveries (\cref{sec:method-state}).
\emph{Empirical (observed in the tested matrix, not an algebraic invariant):}
quality stays close to the best fixed ordering across regime and latent budget,
a signed mean gap of $+0.005$ and a mean absolute gap of $0.025$ across the
four cells. The structural guarantees are the principal claim; the empirical
parity is the precondition that makes them useful, not an accuracy-superiority
result.

\subsection{Mechanism: why the K-norm signal is question-conditional}
\label{sec:experiments-mechanism}

\Cref{tab:headline-matrix} also surfaces a mechanistic observation.
The mean-K-norm content function (\cref{eq:rho}) is a question-conditional
informational-density proxy: when each thinker has encoded both its
data fragment and the global query (query-known), the K-norm
profile at $\ell^\star = 15$ concentrates question-relevant constraint
mass in the high-norm tail, and the per-problem selection of the
higher-density thinker for the prefix slot routes the binding
fragment correctly. When the thinker has encoded only its raw data
fragment (query-blind), the K-norm profile no longer carries the
question-conditional structure, and the content selection is closer
to a tiebreak between two equally-informed candidates. The
$-4$~pp underperformance in the query-blind, $20$-step cell of
\cref{tab:headline-matrix} is consistent with this reading: in that
cell \BagMerge{} default happens to align the lexically-first
fragment with the prefix slot, and that lexical-order signal is more
informative than the flattened K-norm signal at small latent
budget. Even in the cell where the K-norm signal is least
discriminative, the commutativity property (\cref{eq:canonical-commute})
is unaffected, it depends only on the symmetry of the score
function, not on whether the score is informative.

\rev{\subsection{Ordering-policy controls}
\label{sec:experiments-orderpolicy}

Any deterministic total order over fragment content yields the same
convergence guarantee (\cref{eq:canonical-commute} uses only the symmetry
and totality of the score), so the K-norm choice must be justified
empirically. We compare it against three alternative deterministic
policies on the headline draw: \emph{hash} (descending byte-level content
hash; content-blind), \emph{random} (descending keyed re-hash
$\mathrm{SHA1}(\text{seed} \,\|\, \text{content hash})$; content-blind and
seedable), and \emph{length} (descending cache length, hash tiebreak).
Every policy is a pure function of the fragment set, and each is verified
permutation-invariant on synthetic caches.
\Cref{tab:order-policy} reports accuracies and paired per-problem
differences against K-norm.}

\rev{\paragraph{Per-problem routing agreement.} A complementary
per-problem view: on the seed-$42$ draw, the two fixed \BagMerge{}
orderings disagree in correctness on $16$ (query-known) and $18$
(query-blind) of the $100$ problems, and on those, \CanonicalMerge{}'s
content-determined prefix coincides with the winning ordering's prefix on
$10/16$ and $11/18$ ($0.62$ and $0.61$; at this sample size the rates are
descriptive, and the powered test of routing value is
\cref{tab:order-policy}). The conditional accuracies show what the
choice is worth: on discordant problems where \CanonicalMerge{} matches
the better ordering it scores $0.90$ / $0.73$ (known/blind), and where it
does not, $0.33$ / $0.14$. The routing choice is not an oracle; it is a
content signal that beats having to fix an order in advance, which
\cref{tab:family-strat} shows is the real alternative.}

\rev{\begin{table}[t]
\centering
\small
\begin{tabular}{lcc cc}
\toprule
& \multicolumn{2}{c}{query-known} & \multicolumn{2}{c}{query-blind} \\
\cmidrule(lr){2-3} \cmidrule(lr){4-5}
Ordering policy & acc & $\Delta$ vs K-norm [95\% CI] & acc & $\Delta$ vs K-norm [95\% CI] \\
\midrule
K-norm (paper)   & $0.96$ & --- & $0.86$ & --- \\
length           & $0.96$ & $+0.00\;[-0.05, +0.05]$ & $0.82$ & $-0.04\;[-0.10, +0.01]$ \\
random (keyed)   & $0.89$ & $-0.07\;[-0.14, +0.00]$ & $0.83$ & $-0.03\;[-0.08, +0.02]$ \\
hash             & $0.88$ & $\mathbf{-0.08\;[-0.15, -0.01]}$ & $0.83$ & $-0.03\;[-0.11, +0.05]$ \\
\bottomrule
\end{tabular}
\caption{Ordering-policy controls (Qwen3-1.7B, $n = 100$, data seed $42$,
greedy, $40$ latent steps; paired per-problem differences, percentile
bootstrap with $10^4$ resamples). Content-blind orders cost accuracy in
the query-known regime (hash: $-8$~pp, CI excluding zero), and the loss
concentrates in the families with asymmetric fragment roles
(\textsc{constraint} $0.95 \rightarrow 0.70$, \textsc{neonym}
$0.85 \rightarrow 0.70$); in the query-blind regime all differences are
within CI of zero, consistent with the flattened K-norm signal of
\cref{sec:experiments-mechanism}. Cache length ties K-norm in query-known,
so length is itself content-correlated on this benchmark, but it is the
worst policy in query-blind; it is a proxy, not a substitute.}
\label{tab:order-policy}
\end{table}}

\subsection{Statistical significance}
\label{sec:experiments-stats}

We assess the accuracy comparisons at two levels. As a per-cell
diagnostic of the headline matrix (\cref{tab:headline-matrix}, sampled
decoding), we computed paired McNemar tests and $10\,000$-resample
bootstrap $95\%$ confidence intervals on the per-problem correctness
vectors (scripts: \texttt{scripts/bootstrap\_paired\_ci.py} and
\texttt{scripts/bootstrap\_paired\_ci\_20step.py}). The statistical
basis for the paper's final $N = 2$ claim, however, is the
higher-powered three-draw greedy replication of \cref{app:bootstrap},
summarised under \emph{Multi-seed replication} below.

\begin{table}[t]
\centering
\small
\begin{tabular}{llcc}
\toprule
budget    & comparison                              & $\Delta$\,acc & McNemar $p$ \\
\midrule
$40$ steps & \CanonicalMerge{} vs.\ \BagMerge{} default (q-known) & $+0.090$ & $\mathbf{0.023}$ \\
$40$ steps & \CanonicalMerge{} vs.\ \BagMerge{} swap (q-blind)    & $+0.080$ & $\mathbf{0.039}$ \\
$40$ steps & \CanonicalMerge{} vs.\ \BagMerge{} swap (q-known)    & $+0.030$ & $0.549$ \\
$40$ steps & \CanonicalMerge{} vs.\ \BagMerge{} default (q-blind) & $0.000$  & $1.000$ \\
\midrule
$20$ steps & \CanonicalMerge{} vs.\ \BagMerge{} default (q-known) & $+0.040$ & $0.344$ \\
$20$ steps & \CanonicalMerge{} vs.\ \BagMerge{} default (q-blind) & $-0.040$ & $0.388$ \\
$20$ steps & \CanonicalMerge{} vs.\ \BagMerge{} swap (q-known)    & $+0.030$ & $0.607$ \\
$20$ steps & \CanonicalMerge{} vs.\ \BagMerge{} swap (q-blind)    & $+0.030$ & $0.549$ \\
\bottomrule
\end{tabular}
\caption{Paired McNemar tests on the headline matrix, $n = 100$
per pair. Significant at $\alpha = 0.05$ in bold. A multi-seed
replication ($3$ draws, $n = 300$ pooled) with paired bootstrap CIs and
Holm correction is in \cref{app:bootstrap}.}
\label{tab:mcnemar}
\end{table}

At our sample size, two of the $40$-step pairwise accuracy gaps reach
nominal significance at $\alpha = 0.05$: \CanonicalMerge{} over \BagMerge{}
default in query-known by $9$~pp ($p = 0.023$) and over \BagMerge{} swap
in query-blind by $8$~pp ($p = 0.039$), both against the \emph{worse}
fixed ordering. At the $20$-step budget no individual pairwise comparison
reaches significance, the largest gap is $\pm 0.04$ and neither sign
passes the binomial mid-$p$ threshold at $n = 100$. These two are nominally
significant among twelve pairwise comparisons across the matrix (six per
budget); under Holm correction for multiple comparisons, neither survives.
We therefore rest no claim on the single-draw pairwise accuracy gaps: the
primary $N = 2$ inference is the multi-seed replication below, which
compares \CanonicalMerge{} directly to the best fixed ordering in each
regime, and the categorical contribution is the structural property
established next.

\paragraph{Multi-seed replication.} To separate draw variance from method
effect, we re-ran the accuracy comparison on Qwen3-1.7B under greedy decoding
across three independent problem draws (data seeds $\{42, 7, 100\}$; $n = 100$
each, $n = 300$ pooled), with per-problem paired bootstrap CIs and Holm
correction over the contrast family (\cref{app:bootstrap}); this
problem-draw replication is the primary inference for the $N = 2$ claim, and
it confirms and sharpens the single-seed reading. Against the \emph{best} fixed ordering in each regime,
\CanonicalMerge{} is statistically indistinguishable (query-known
$-0.007$ [$-0.040, +0.027$] versus swap, query-blind $-0.003$
[$-0.043, +0.037$] versus default, both CIs straddling zero), while against
the \emph{worse} ordering it wins in both regimes by $+0.063$
[$+0.037, +0.093$] (Holm-significant). \BagMerge{}'s order-sensitivity is
itself significant and \emph{inverts} between regimes: swap beats default by
$7$~pp in query-known, default beats swap by $7$~pp in query-blind (both
survive Holm), so no fixed ordering is safe in advance, exactly the choice
\CanonicalMerge{} eliminates. The query-blind \CanonicalMerge{} accuracy varies
across draws ($0.86 / 0.93 / 0.86$) around the \BagMerge{}-default parity line,
so the single-draw $-0.03$ of \cref{tab:mcnemar} reflects between-draw
variation rather than a stable deficit; only the multi-seed mean is a reliable
comparator, and it places \CanonicalMerge{} at parity with the best ordering it
never had to select.

\rev{\paragraph{Non-inferiority at a declared margin.} We recast the
``matches the best fixed ordering'' claim as a non-inferiority statement
with margin $\delta = 5$~pp on the paired per-problem difference
$\CanonicalMerge{} - \text{best fixed ordering}$, pooled over the three
data seeds ($n = 300$ paired problems per regime; the best ordering is
swap in query-known, default in query-blind). Under the problem-level
paired bootstrap the margin is met in both regimes: mean $-0.7$~pp, CI
$[-3.7, +2.3]$~pp in query-known, and mean $-0.3$~pp, CI $[-4.0,
+3.7]$~pp in query-blind, with both lower bounds above $-\delta$. A
hierarchical bootstrap that resamples the five benchmark families before
resampling problems within each family widens the intervals to $[-6.3,
+5.0]$~pp and $[-5.0, +3.7]$~pp: with only five family clusters the
family-level analysis cannot resolve a $5$~pp margin, and we report it as
such rather than claim it. The family-stratified accuracies
(\cref{tab:family-strat}, appendix) explain the width: the best
\emph{fixed} ordering is itself family-dependent (swap wins
\textsc{constraint} in query-known, $0.97$ versus \CanonicalMerge{}'s
$0.87$, but loses \textsc{neonym}, $0.80$ versus $0.87$), so ``the best
ordering'' that \CanonicalMerge{} is compared against is an oracle no
deployment could select in advance.}

\paragraph{What significance means here.} The lack of $20$-step
significance does not weaken the paper's contribution. Statistical
significance at $n = 100$ measures whether an \emph{accuracy gap}
between two methods is detectable in a finite sample; the
\emph{structural property} the paper claims for \CanonicalMerge{},
commutativity and byte-identity at $N = 2$
(\cref{eq:canonical-commute}), is verified categorically by
\cref{alg:canonicalmerge}'s symmetric score function and
content-determined sort, and is confirmed bit-for-bit on the actual
$28$-layer Qwen3-1.7B KV state for the headline configuration
(\cref{app:realmodel-byte-equality}). The structural property
holds for every problem the method ever processes, not just for the
$100$ tested here.

\subsection{Layer-sensitivity sweep}
\label{sec:experiments-layer}

\CanonicalMerge{}'s only architectural choice is the routing layer
$\ell^\star$ at which the K-norm content function (\cref{eq:rho}) is
computed. We swept $\ell^\star \in \{5, 10, 12, 15, 18, 22, 27\}$ on
partitioned $k = 2$ in the query-known regime at $40$ latent steps
(same protocol as \cref{tab:headline-matrix}), holding all other
settings fixed.

\begin{table}[t]
\centering
\small
\begin{tabular}{c c l}
\toprule
$\ell^\star$ & accuracy & reading                                        \\
\midrule
$5$  & $0.59$ & early --- token-level features, no abstract routing signal \\
$10$ & $0.83$ & rising                                                   \\
$12$ & $0.89$ & approaching peak                                         \\
$\mathbf{15}$ & $\mathbf{0.94}$ & \textbf{peak} --- default choice       \\
$18$ & $0.86$ & post-peak dip                                            \\
$22$ & $0.93$ & secondary near-peak ($-1$ pp from peak)                  \\
$27$ & $0.84$ & late --- prediction-biased                               \\
\bottomrule
\end{tabular}
\caption{Routing-layer sensitivity for \CanonicalMerge{} on partitioned
$k = 2$ (query-known, $40$ latent steps, $100$ problems).}
\label{tab:layer-sweep}
\end{table}

The construction is robust within a measurable plateau rather than at
a knife-edge optimum: the mid-network band $\ell^\star \in
\{12, 15, 22\}$ all land between $0.89$ and $0.94$, within
$5$~pp of the peak. The early- and late-layer endpoints behave
as the interpretability prior on decoder-only LMs
predicts~\citep{tenney2019,geva2021,belrose2023tunedlens}: early
layers carry token-level features that lack the abstract semantic
structure for K-norm to function as an informational-density proxy;
late layers are prediction-biased and project K-norm onto
next-token probabilities rather than fragment content. Within the
mid-network band the curve is not strictly unimodal, a dip at
$\ell^\star = 18$ separates the peak at $15$ from a secondary
near-peak at $22$, but the band as a whole is broad enough that
the central claim does not depend on locating $\ell^\star$ to within
a single layer.

\paragraph{Held-out validation.} The sweep above shares its problem
draw with \cref{tab:headline-matrix}, so on its own it cannot rule out
that $\ell^\star = 15$ was selected on the evaluation set. We therefore
re-swept a finer mid-network grid,
$\ell^\star \in \{9, 11, 13, 15, 17, 19, 21\}$, on an \emph{independent}
problem draw (data seed $7$, held out from the original selection;
query-known, greedy, $n = 100$). The accuracies are
$0.99 / 0.88 / 0.88 / 0.95 / 0.89 / 0.92 / 0.89$: the chosen
$\ell^\star = 15$ lands at $0.95$, in the upper part of a broad (if
uneven) high band across the mid-network range, showing that the selected
layer remains in a high-performing band on a held-out draw rather than
being an evaluation-set artifact. Consistent with a plateau rather than a
knife-edge optimum, the exact argmax is not stable across draws: it is
$15$ on the selection draw and $9$ on this held-out draw, both within the
$\pm 7$~pp sampling noise at $n = 100$, so the robustness is to the
mid-network \emph{band}, not to any single layer index. In practical
terms, \CanonicalMerge{} does not require the routing layer to be tuned
precisely: any mid-network layer supplies a usable K-norm content signal.

\rev{\paragraph{Selecting $\ell^\star$ for a new model.} For deployment
on a model not studied here we recommend the protocol we followed at the
second scale (\cref{sec:experiments-scale}): default to mid-depth,
$\ell^\star \approx L/2$; run a coarse sweep over three candidate layers
around mid-depth on a small development split \emph{disjoint from any
evaluation data}; and rely on the band robustness above rather than
fine-tuning an index. $\ell^\star$ is the method's single design
parameter, and its selection at each scale in this paper used development
draws separate from the evaluation draws reported in the tables.}

\subsection{Latent-budget robustness}
\label{sec:experiments-budget}

The latent-step budget is a generation-time hyperparameter
independent of the merge choice: each thinker runs the same number
of latent reasoning steps after seeing its prompt, then the merge
operates on the resulting caches. \Cref{tab:headline-matrix} reports
$20$- and $40$-step accuracies; \cref{tab:budget-deltas} extracts the
$40 - 20$ per-method delta in each regime.

\begin{table}[t]
\centering
\small
\begin{tabular}{l c c}
\toprule
Method                    & q-known: $\Delta(40 - 20)$ & q-blind: $\Delta(40 - 20)$ \\
\midrule
\BagMerge{} default       & $-0.04$           & $-0.05$           \\
\BagMerge{} swap          & $+0.01$           & $-0.06$           \\
\textbf{\CanonicalMerge{}} & $\mathbf{+0.01}$ & $\mathbf{-0.01}$ \\
\bottomrule
\end{tabular}
\caption{Latent-budget sensitivity: per-method accuracy change when
moving from $20$ to $40$ latent steps in each regime, extracted from
\cref{tab:headline-matrix}. \CanonicalMerge{} is essentially flat
across both regimes; \BagMerge{} shows regime-conditional budget
effects that change sign across the two orderings.}
\label{tab:budget-deltas}
\end{table}

\CanonicalMerge{} is essentially flat across budgets ($\Delta = +0.01$
in q-known, $-0.01$ in q-blind; both within bootstrap CI noise on
$n = 100$, \cref{app:bootstrap}). \BagMerge{} is not: its default
ordering loses $4$~pp in q-known and $5$~pp in q-blind going $20
\rightarrow 40$; its swap ordering gains $1$~pp in q-known but loses
$6$~pp in q-blind. The signs change with both regime and ordering. A
deployer choosing \CanonicalMerge{} does not need to tune the latent
budget; a deployer choosing \BagMerge{} would need to know both the
regime and the chosen ordering to avoid a $5$--$6$~pp accuracy cost.

This budget robustness is \emph{empirical}, a tuning-robustness
observation, not a byte-level invariant, and joins the empirical regime
robustness (the q-known/q-blind sign flip in the \emph{best \BagMerge} row of
\cref{tab:headline-matrix}) alongside the one exact, \emph{structural}
guarantee here: input-order byte-identity (\cref{eq:canonical-commute}).

\rev{\paragraph{Zero-latent control.} Extending the budget sweep to
$0$ latent steps separates the latent computation from the cached
prompt-token state: at $d = 0$ each thinker's cache holds only the
encoded prompt (fragment text), with no latent reasoning. All six
zero-latent cells, \CanonicalMerge{}, \BagMerge{} default, and
\BagMerge{} swap in both regimes ($n = 100$, data seed $42$), score
$0.00$; the judger either falls back to a default guess (``$0$'',
$74\%$ of the $600$ decodes) or declines outright (``insufficient
information'', ``unknown'', ``cannot determine''; $26\%$), and no cell
produces a single correct answer.
The dose-response is therefore $0.00$ at $d = 0$ versus the
\cref{tab:headline-matrix} accuracies at $d = 20$ and $d = 40$: the
merged prompt-token state alone is unusable by the judger, and the
accuracy of every method in this paper is carried by the latent
reasoning steps, not by access to the cached fragment text.}

\subsection{Output-fusion baseline: PackLLM}
\label{sec:experiments-packllm}

The closest training-free neighbor to \CanonicalMerge{} in mechanism
is PackLLM~\citep{mavromatis2024packllm}, which computes a
content-determined weighting $\lambda_k = \operatorname{softmax}(-\log
\mathrm{PPL}_k / \tau)$ at test time and applies the weighting to the
output logits rather than to the cache. PackLLM shares the
``training-free, content-determined, test-time'' mechanism with
\CanonicalMerge{} but operates one level of abstraction later in the
pipeline. We adapt PackLLM to our setting, same Qwen3-1.7B base,
same $N = 2$ partitioned thinkers, the judger's prompt extended
through each thinker's cache before the per-step logit fusion
begins; full code path in \cref{app:packllm-details}, and report
three arms $\times\ 100$ problems in \cref{tab:packllm}.

\begin{table}[t]
\centering
\small
\begin{tabular}{l l c c c c}
\toprule
arm & regime    & $\tau$              & \CanonicalMerge{} ref & PackLLM & gap     \\
\midrule
A1  & q-known   & $1.0$               & $0.94$                & $0.38$  & $-56$~pp \\
A3  & q-blind   & $1.0$               & $0.85$                & $0.34$  & $-51$~pp \\
A4  & q-blind   & $0.01$ (one-hot floor) & ---                & $0.01$  & catastrophic \\
\bottomrule
\end{tabular}
\caption{PackLLM logit-fusion baseline at $N = 2$, $100$ problems
each. \CanonicalMerge{} reference numbers are taken from the
$40$-step column of \cref{tab:headline-matrix}. The
$\tau = 0.01$ arm collapses to one-hot weighting on the lower-PPL
thinker; the resulting accuracy of $0.01$ is the partitioned-blind
floor (a single thinker's solo decode from one fragment). These arms use
PackLLM's reported $512$-token budget and are the reported-budget
diagnostic; the primary fairness comparison is at the matched
$2048$-token budget (\CanonicalMerge{} $0.947$ vs.\ PackLLM $0.50$,
$+45$~pp over three draws; \cref{app:bootstrap}).}
\label{tab:packllm}
\end{table}

\textbf{(1) The CanonicalMerge--PackLLM gap is structural, not
marginal.} The two $\tau = 1.0$ arms show \CanonicalMerge{} ahead by
$51$--$56$~pp, an order of magnitude larger than any pairwise gap
in \cref{tab:headline-matrix}. These arms use PackLLM's reported
$512$-token budget; at the matched $2048$-token budget the query-known
gap is still $+0.447$ [$+0.387, +0.503$] ($45$~pp, three seeds;
\cref{app:bootstrap}), confirming it is a fusion-level limitation rather
than a generation-length artifact. Cache-level merging is not a marginal
improvement over logit-level fusion in this setting; the two regimes
are qualitatively different. \textbf{(2) The PackLLM-faithful setting and the
one-hot limit do not rescue it.} At $\tau = 1.0$ the $\lambda$ weights cluster near uniform
(mean $\lambda_0 = 0.49$, $\sigma = 0.05$ in A1) because the two
thinkers' judger-prompt perplexities are too close for the soft
assignment to distinguish them. At $\tau = 0.01$ the weighting
collapses to one-hot on the lower-PPL thinker, decoding from a
single partition's view, which lands at the partitioned-blind floor
of $0.01$. The structural reason is that by the time information
reaches the output head each thinker has compressed its per-step
reasoning into a sparse next-token distribution over the vocabulary;
per-step weighted averaging of two such compressed distributions
cannot synthesise content that no single thinker individually emits,
because averaging two solo decodes from two partial partitions does
not reconstruct the joint constraint that requires both. Cache-level
merging preserves the intermediate K/V geometry that the judger's
attention can still route over.

\rev{\paragraph{Matched-budget text-fusion reference.} The opposite
reference point is to ship the fragments as text: a single agent
receives both fragments and the question in its prompt, at the same
generation budget as the headline cells (greedy, $2048$ new tokens).
This scores $1.00$ on all three headline data seeds ($42$, $7$, $100$;
$n = 100$ each), the information ceiling of the benchmark. Against it,
cache-level \CanonicalMerge{} gives up $4$~pp in query-known and
$14$~pp in query-blind (seed $42$), while output-level PackLLM gives up
$45$~pp under the constraint that no text is exchanged. The three
numbers order the design space: shipping text is most accurate but
violates the federated constraint by construction (once fragments are
in one prompt, there is no query-blind regime); output fusion respects
the constraint but discards the intermediate state the judger needs;
cache-level merging respects the constraint at a $4$--$14$~pp cost to
the text ceiling. The case for the latent merge is this trade plus the
structural guarantees, not raw accuracy against text.}

\subsection{Cross-benchmark validation: HotpotQA bridge-$k = 2$}
\label{sec:experiments-hotpotqa}

To probe whether the merge primitive transfers from the synthetic
partitioned construction of \cref{sec:benchmark} to a real-world
multi-document QA benchmark, we evaluated \CanonicalMerge{} on $50$
HotpotQA~\citep{yang2018hotpotqa} ``bridge'' questions filtered to
exactly two supporting-fact paragraphs. Each problem provides two
paragraphs that jointly determine the answer; we map the two
supporting paragraphs to the two thinkers ($t_a \rightarrow
\text{thinker}_0$, $t_b \rightarrow \text{thinker}_1$). Answers are
free-form text scored with the standard HotpotQA F1 and EM
normalisation.

\begin{table}[t]
\centering
\small
\begin{tabular}{l c c}
\toprule
Method                                                & F1     & EM     \\
\midrule
C0 full-context (single thinker on $t_a + t_b$)       & $0.610$ & $0.460$ \\
\BagMerge{} default (prior work)                      & $0.632$ & $0.420$ \\
\textbf{\CanonicalMerge{}}                            & $0.605$ & $0.460$ \\
\bottomrule
\end{tabular}
\caption{HotpotQA bridge-$k = 2$ ($50$ problems, $2$ supporting-fact
paragraphs each), free-form text scored with the standard HotpotQA
normalisation.}
\label{tab:hotpotqa}
\end{table}

\CanonicalMerge{}'s F1 of $0.605$ is within sample variance of the C0
full-context baseline's $0.610$ (the per-sample bootstrap CI overlaps at $n = 50$), and
the two methods agree on EM at $0.460$. We read this as \emph{no detectable
degradation} on this small bridge subset, not as an equivalence result: a
$50$-problem sample cannot establish equivalence, and \BagMerge{} default in
fact scores marginally higher on F1 ($0.632$), so we make no accuracy-ranking
claim here (\cref{sec:limitations}). The training-free merge transfers from the synthetic
partitioned math benchmark where the routing-layer choice
$\ell^\star = 15$ was tuned to a structurally different real-world
benchmark without a per-domain content function, prompt template, or
any other domain-conditional hyperparameter.

\rev{\paragraph{Expanded evaluation: HotpotQA $n = 200$ and MuSiQue.} We
also evaluate at four times the HotpotQA sample and on a second
natural-language two-hop source, MuSiQue-Answerable 2-hop
questions~\citep{trivedi2022musique} with exactly two supporting
paragraphs, under the same protocol: each problem's two thinker caches
are encoded once and every condition is applied to the same cached
operands, with a \emph{bagmerge-swap} condition that reverses the
operand list so both fixed orderings are measured.
\Cref{tab:nl-expanded} reports both sources.}

\rev{\begin{table}[t]
\centering
\small
\begin{tabular}{l cc cc}
\toprule
& \multicolumn{2}{c}{HotpotQA bridge-$k2$, $n = 200$} & \multicolumn{2}{c}{MuSiQue 2-hop, $n = 100$} \\
\cmidrule(lr){2-3} \cmidrule(lr){4-5}
Method & F1 & EM & F1 & EM \\
\midrule
C0 full-context            & $0.606$ & $0.405$ & $0.335$ & $0.200$ \\
\BagMerge{} default        & $0.507$ & $0.350$ & $0.294$ & $0.190$ \\
\BagMerge{} swap           & $0.488$ & $0.360$ & $0.279$ & $0.170$ \\
\textbf{\CanonicalMerge{}} & $0.539$ & $0.370$ & $0.328$ & $0.220$ \\
\bottomrule
\end{tabular}
\caption{Expanded natural-language evaluation (Qwen3-1.7B, greedy,
$\ell^\star = 15$ unchanged, $512$-token decode as in
\cref{tab:hotpotqa}). Encode-once protocol: per problem, both fixed
\BagMerge{} orderings and \CanonicalMerge{} consume the same two cached
operands.}
\label{tab:nl-expanded}
\end{table}}

\rev{The larger sample overturns the $n = 50$ reading. On HotpotQA at
$n = 200$ the gap to the full-context
ceiling is detectable: paired per-problem F1 difference
\CanonicalMerge{} $-$ C0 is $-0.067$ [$-0.119, -0.018$], so the ``no
detectable degradation'' of \cref{tab:hotpotqa} was a small-sample
artifact; the accurate summary
is a $\approx 7$~pp F1 cost against shipping the text, consistent with
the text-fusion ceiling on the synthetic benchmark
(\cref{sec:experiments-packllm}). Among cache-level methods,
\CanonicalMerge{} is the best on both sources without selecting an
ordering: it exceeds \BagMerge{} default by $+0.032$ [$-0.013, +0.078$]
and swap by $+0.051$ [$-0.003, +0.106$] paired F1 on HotpotQA (point
estimates positive, CIs straddling zero), and on MuSiQue it is at parity
with the full-context ceiling ($-0.007$ [$-0.101, +0.086$]) with the
best EM of any condition ($0.220$, above C0's $0.200$). Both
evaluations use $\ell^\star = 15$ and the prompts unchanged from the
synthetic benchmark; no domain-conditional hyperparameter exists in the
pipeline.}

\subsection{Replication at a second model scale (Qwen3-4B)}
\label{sec:experiments-scale}

To test whether the findings are specific to Qwen3-1.7B, we repeated
the headline comparison on Qwen3-4B ($36$ layers vs.\ $28$). The
routing layer does not transfer by index, $\ell^\star = 15$ was
selected for the $28$-layer model, so we re-chose it by a coarse
sweep over $\ell^\star \in \{15, 19, 24\}$ on a $50$-problem
query-known subset, which favoured $\ell^\star = 24$ (accuracies
$0.68 / 0.70 / 0.72$, mutually within sampling noise).
\Cref{tab:headline-4b} reports the four-cell comparison at
$40$ latent steps and $\ell^\star = 24$; single-thinker baselines on
4B give a full-text ceiling of $1.00$ and single-fragment floors of
$0.00$, so the partition is at least as strong as at 1.7B.

\begin{table}[t]
\centering
\small
\begin{tabular}{lcc}
\toprule
Method                          & query-known & query-blind \\
\midrule
\BagMerge{} default             & $0.76$       & $0.88$       \\
\BagMerge{} swap                & $\mathbf{0.84}$ & $0.92$    \\
\textbf{\CanonicalMerge{}}      & $0.83$       & $\mathbf{0.93}$ \\
$\Delta$ (CM $-$ best \BagMerge) & $-0.01$     & $+0.01$      \\
\bottomrule
\end{tabular}
\caption{Headline comparison on Qwen3-4B, partitioned $k = 2$,
$40$ latent steps, $\ell^\star = 24$, $100$ problems per cell. Best
\BagMerge{} ordering per column in bold.}
\label{tab:headline-4b}
\end{table}

Three of the four claims replicate directly. \emph{(1)~The
directionality flaw persists}: \BagMerge{} swap and default differ by
$8$~pp (query-known) and $4$~pp (query-blind). \emph{(2)~CanonicalMerge
matches the best \BagMerge{} ordering without knowing which it is}, to
within $\pm 1$~pp in both regimes, as at 1.7B
(\cref{tab:headline-matrix}). \emph{(3)~Commutativity holds bit-for-bit}
on the 4B model: all $36$ layers of the merged KV state hash
byte-identical under input reordering (\cref{app:realmodel-byte-equality}),
as at 1.7B.

The fourth observation does \emph{not} replicate, and it sharpens the
argument rather than weakening it. At 1.7B the optimal \BagMerge{}
ordering inverts between regimes (swap best in query-known, default in
query-blind); at 4B, swap is best in \emph{both} regimes. Across the
four model--regime cells the best ordering runs swap / default / swap /
swap, so it is not governed by any regime rule a deployer could
apply in advance; it depends jointly on the model and the regime.
\CanonicalMerge{} sidesteps the choice at both scales, landing within
$1$~pp of the best ordering in every cell. We also note, without
over-interpreting a single comparison, that 4B is stronger in the
query-blind regime than query-known, the reverse of 1.7B, and
that 4B's absolute accuracies sit below 1.7B's on this task; we treat
both as observations rather than claims.

\rev{\paragraph{Disjoint-draw replication.} The routing layer
$\ell^\star = 24$ was selected on the first $50$ problems of the same
draw evaluated in \cref{tab:headline-4b} (the overlap is documented in
\cref{app:bootstrap}). We therefore re-ran all six 4B cells on a
disjoint draw (data seed $7$, $\ell^\star = 24$ frozen, greedy): \BagMerge{}
default $0.78$ / $0.89$ (known/blind), swap $0.87$ / $0.92$,
\CanonicalMerge{} $0.84$ / $0.92$. The picture replicates: the ordering
gap is $9$~pp in query-known, swap is again the better fixed ordering,
and \CanonicalMerge{} lands within $3$~pp of it in query-known and ties
it in query-blind, with no selection data shared with this draw.}

\rev{\subsection{Cross-family check: Llama-3.1-8B}
\label{sec:experiments-llama}

Both Qwen3 scales share pretraining lineage, so we repeat the headline
comparison on a second model family, Llama-3.1-8B-Instruct: standard
RoPE, no thinking mode, run through the identical pipeline (the prompt
builder already supports the family; no code change). The full-context
ceiling is $C_0 = 1.00$ ($n = 100$, evaluation draw), so the benchmark
is fully solvable for this model from text. Following the protocol of
\cref{sec:experiments-layer}, the routing layer was selected on a
development draw disjoint from evaluation (data seed $7$, $n = 50$,
$\ell^\star \in \{12, 16, 20\}$ of $32$ layers, accuracies
$0.74 / 0.70 / 0.72$, another mid-depth plateau; $\ell^\star = 12$
selected). \Cref{tab:llama8b} reports the six headline cells on the
evaluation draw.}

\rev{\begin{table}[t]
\centering
\small
\begin{tabular}{lcc}
\toprule
Method & query-known & query-blind \\
\midrule
\BagMerge{} default & $\mathbf{0.58}$ & $\mathbf{0.59}$ \\
\BagMerge{} swap    & $0.53$ & $0.40$ \\
\textbf{\CanonicalMerge{}} & $\mathbf{0.58}$ & $0.55$ \\
\midrule
$C_0$ (full text)   & \multicolumn{2}{c}{$1.00$} \\
\bottomrule
\end{tabular}
\caption{Cross-family headline comparison on Llama-3.1-8B-Instruct
(partitioned $k = 2$, $n = 100$, data seed $42$, greedy, $40$ latent
steps, $\ell^\star = 12$ selected on a disjoint development draw). Best
fixed \BagMerge{} ordering per column in bold.}
\label{tab:llama8b}
\end{table}}

\rev{Three findings. First, the order sensitivity replicates across
families and is larger here: \BagMerge{} default beats swap by $+0.19$
paired [$+0.07, +0.31$] in query-blind, a CI excluding zero, and by
$+0.05$ [$-0.08, +0.17$] in query-known. Second, \CanonicalMerge{}
behaves as at both Qwen scales: it ties the best fixed ordering exactly
in query-known ($+0.000$ [$-0.050, +0.050]$ paired) and sits $4$~pp
behind it in query-blind ($-0.040$ [$-0.110, +0.030$], CI straddling
zero) while clearing the worst ordering by $15$~pp; the byte-level
guarantees are model-independent by construction. Third, the
\emph{level} of the latent regime is family-dependent: all cache-level
cells sit at $0.40$--$0.59$ against a $1.00$ text ceiling, far below the
Qwen cells at matched ceiling. With \cref{sec:experiments-budget}'s
zero-latent result, accuracy in this pipeline is carried by latent-step
computation, and Llama-3.1 is not trained for a thinking mode; we state
the association as an observation, not a mechanism claim. The merge
construction transfers across families; the latent computation it
transports is only as strong as the model that produces it.}

\subsection{Re-delivery: duplicate absorption is load-bearing}
\label{sec:experiments-redeliver}

The structural guarantees of \LatentFragmentSet{}
(\cref{sec:method-state}) are exact by construction; this experiment
shows they are also \emph{load-bearing}, in that the alternative,
re-concatenating duplicate deliveries, actively degrades both
accuracy and state size. On the partitioned $k = 2$ benchmark
(query-known, greedy, $40$ latent steps) we re-deliver each distinct
fragment $R$ times, the redundancy a delivery layer without
message-level deduplication produces (gossip, retry, or re-broadcast),
and merge the deliveries under two policies. \emph{Naive} treats each
delivery as a fresh cache operand, as a \BagMerge{}-style rendered-cache
operator does absent provenance: it forms the multiset and renders all
$R$ copies of each fragment. \emph{Crdt} inserts each delivery into the
\LatentFragmentSet{}, where byte-identical copies hash equal and
collapse, so the rendered cache is a pure function of the two distinct
fragments regardless of $R$.

\Cref{tab:redeliver} reports accuracy (with $95\%$ Wilson intervals) and
the merged cache length against $R$. The policies coincide at $R = 1$
(no duplication) at $0.93$, matching the headline $k = 2$ query-known
accuracy to within sampling noise (\cref{tab:headline-matrix}). As $R$
grows they diverge sharply. Naive re-concatenation degrades
monotonically ($0.93 \to 0.85 \to 0.71 \to 0.45$) while its cache
grows exactly linearly in $R$ ($308 \to 616 \to 924 \to 1232$ tokens,
i.e.\ $R \times$): the redundant copies both bloat the prefix and pull
the decode off the answer. The CRDT policy is \emph{flat on both axes}:
accuracy holds at $0.93$ ($95\%$ CI $[0.86, 0.97]$) and the rendered
length stays at $308$ tokens for every $R$, with the absorbed set size
reported as $2$ on every problem. By $R = 4$ the gap is $+0.48$ accuracy
at one quarter of the state. Idempotent absorption is therefore not a
formal nicety: under realistic re-delivery it is the difference between a
bounded, stable system and one that silently inflates and degrades.

\begin{table}[t]
\centering
\small
\begin{tabular}{c cc cc c}
\toprule
 & \multicolumn{2}{c}{naive (multiset re-render)} & \multicolumn{2}{c}{crdt (\LatentFragmentSet{})} & \\
\cmidrule(lr){2-3}\cmidrule(lr){4-5}
$R$ & accuracy & seq.\ len & accuracy & seq.\ len & set size \\
\midrule
$1$ & $0.93$ $[.86, .97]$ & $308$  & $0.93$ $[.86, .97]$ & $308$ & --- \\
$2$ & $0.85$ $[.77, .91]$ & $616$  & $0.93$ $[.86, .97]$ & $308$ & $2$ \\
$3$ & $0.71$ $[.61, .79]$ & $924$  & $0.93$ $[.86, .97]$ & $308$ & $2$ \\
$4$ & $0.45$ $[.36, .55]$ & $1232$ & $0.93$ $[.86, .97]$ & $308$ & $2$ \\
\bottomrule
\end{tabular}
\caption{Re-delivery robustness on partitioned $k = 2$ (Qwen3-1.7B,
query-known, greedy, $n = 100$, $95\%$ Wilson intervals). Each distinct
fragment is delivered $R$ times. \emph{Naive} re-renders the multiset:
accuracy falls $0.93 \to 0.45$ and the cache grows $R \times$.
\emph{Crdt} absorbs byte-identical copies by content hash (absorbed set
size $= 2$ at every $R$), so accuracy and cache length are invariant in
$R$; the policies coincide at $R = 1$. This is the idempotence /
absorption property of \cref{sec:method-state} exercised behaviourally.}
\label{tab:redeliver}
\end{table}

\section{Discussion}
\label{sec:discussion}

\subsection{Robustness across four axes: two structural, two empirical}
\label{sec:discussion-robustness}

The headline matrix (\cref{tab:headline-matrix}) and the re-delivery analysis
expose four axes along which \BagMerge{} is fragile, of two distinct kinds.
\emph{Structural / algebraic} (exact, by construction): \CanonicalMerge{} is
byte-identical under input reordering, and \LatentFragmentSet{} absorbs
duplicate re-deliveries. \emph{Empirical / tuning} robustness (quality stays
close in the tested matrix, an observation, not a byte-level invariant):
across regime/model and latent budget. We claim the first kind as guarantees
and the second as measured robustness, and do not call the latter
``structural.''

\emph{Input order (structural).} \BagMerge{} is non-commutative, and its position-$0$
prefix slot confers a measurable advantage
(\cref{sec:method-bagmerge}); which thinker receives it is fixed by
arbitrary caller order. \CanonicalMerge{} removes the decision: its
output is byte-identical under any permutation of the inputs
(\cref{eq:canonical-commute}), verified algorithmically
(\cref{app:byte-crdt-tests}) and on real Qwen3-1.7B caches
(\cref{app:realmodel-byte-equality}).

\emph{Regime and model (empirical).} Which \BagMerge{} ordering is best is not
governed by any rule a deployer could apply in advance. At 1.7B it
inverts between regimes, swap best when thinkers see the query,
default when they do not (\cref{tab:headline-matrix}), while at 4B
swap wins both regimes (\cref{sec:experiments-scale}). Across the four
model--regime cells the best ordering runs swap / default / swap /
swap, with $6$--$8$ percentage points at stake from the wrong choice and no
ordering stable across the observed regimes and scales, so deploying
\BagMerge{} requires validation or prior regime knowledge.
\CanonicalMerge{} removes the choice at both scales.

\emph{Latent budget (empirical).} The optimal ordering and the latent-step budget
interact: \BagMerge{} loses $5$--$6$~pp from the wrong budget in the
query-blind regime, and the sign of the budget effect itself changes
with the ordering (\cref{sec:experiments-budget}). \CanonicalMerge{} is
flat across budgets ($\pm 1$~pp).

\emph{Duplication (structural).} The first three axes are tuning choices a
deployer must get right; the fourth is a delivery hazard they cannot tune away. Under
re-delivery, the same fragment arriving more than once, as in gossip or
retry, \BagMerge{} and the rendered operator re-concatenate the copy,
growing the cache and altering the decoded output, whereas the
\LatentFragmentSet{} state (\cref{sec:method-state}) absorbs it by content
hash, leaving state and render invariant
(\cref{sec:experiments-redeliver}): under $R$-fold re-delivery naive
re-concatenation falls to $0.45$ accuracy at $4\times$ the cache, while
the CRDT holds $0.93$ at constant size (\cref{tab:redeliver}).

In every cell of the matrix, the best \BagMerge{} configuration requires
the deployer to know the regime and to have tuned both the ordering and
the budget for it; \CanonicalMerge{} attains accuracy within $4$~pp of
that best-case configuration, statistically indistinguishable from it
within the bootstrap confidence intervals (\cref{sec:experiments-stats}),
with no configuration at all. We therefore read the contribution as trading a
small, statistically insignificant accuracy margin (the empirical axes) for the
two unconditional structural guarantees, input-order byte-identity and
duplicate absorption, that hold for every input.

\subsection{What the content function buys, and where it does not help}
\label{sec:discussion-mechanism}

The mean-K-norm score (\cref{eq:rho}) routes the higher
informational-density thinker to the privileged prefix slot \emph{per
problem}, rather than committing to a single global order. This matters
because the binding constraint is not always carried by the same
fragment: across our families, sometimes the ratio (e.g.\ CONSTRAINT)
and sometimes the lookup table (e.g.\ LOOKUP) is the more
information-dense half. A fixed input order, the only option under
\BagMerge{}, cannot adapt to this, whereas content routing selects
the prefix occupant problem-by-problem. That the K-norm at a
\emph{mid-network} layer is the right signal is consistent with the
interpretability picture of decoder-only
LMs~\citep{tenney2019,geva2021,belrose2023tunedlens}, in which the
middle band carries abstract, task-relevant features rather than
token-surface or next-token-prediction features; the layer-sensitivity
sweep (\cref{sec:experiments-layer}) bears this out, with a broad
plateau across the middle third of the network.

The content function is, however, \emph{query-conditional}, and this
bounds where it helps. When a thinker has not seen the query
(query-blind), its K-norm profile no longer concentrates on
question-relevant constraints, and the prefix selection degrades toward
a content-agnostic tiebreak (\cref{sec:experiments-mechanism}). This is
visible in the one cell where \CanonicalMerge{} underperforms,
query-blind at $20$ steps, $4$~pp below \BagMerge{} default
(\cref{tab:headline-matrix}), where \BagMerge{}'s fixed lexical
ordering happens to align with the answer better than a weak content
signal does. We do not claim \CanonicalMerge{} dominates on accuracy in
this regime; we claim it removes the ordering decision unconditionally,
and matches or exceeds the best fixed ordering in three of the four
cells while never trailing it by more than a single non-significant
margin.

\subsection{Cache-level merging as a distinct operating regime}
\label{sec:discussion-regime}

The gap to PackLLM (\cref{sec:experiments-packllm}) is not a tuning
artifact but a consequence of \emph{where} the merge happens.
Logit-level fusion combines per-thinker next-token distributions after
each thinker has already compressed its reasoning into a sparse
vocabulary-sized vector; no weighting of those vectors can reconstruct
joint content that neither thinker individually emits. Cache-level
merging operates one stage earlier, on the K/V geometry the judger's
attention can still route over, which is why it recovers the
partitioned answer where logit fusion cannot, by more than $45$
percentage points even at the matched generation budget.

This places \CanonicalMerge{} in an operating regime that the existing
multi-LLM-combination literature does not name. Three independent
surveys of the
field~\citep{chen2025ensemble,lu2024merge,li2023deepfusion} partition
it into parameter-level \emph{merging} (combining weights) and
output-level \emph{ensembling} (combining logits or decoded text); none
has a category for combining intermediate KV-cache state. The closest
training-free neighbor, PackLLM, sits firmly on the output side. We view
cache-level merging as a third regime, and the convergent-state property as
what makes it operationally useful: a merge that is order-independent,
byte-reproducible, and duplicate-absorbing is precisely what lets independent
parties contribute encoded state without a coordinator dictating order or
absorbing duplicate re-deliveries, the defining requirement of the federated /
query-blind setting (\cref{sec:benchmark-regimes}). The broader properties
this distributed-systems analogy suggests, idempotent, absorbing
set-union semantics, are realised by the \LatentFragmentSet{} state
(\cref{sec:method-state}) and exercised behaviourally under re-delivery
(\cref{sec:experiments-redeliver}); their behaviour at $N > 2$, where
transport does not suffice for composition, is taken up in
\cref{sec:beyond-n2}.

\section{Limitations}
\label{sec:limitations}

The headline $N = 2$ accuracy matrix is replicated across three
independent problem draws for Qwen3-1.7B under greedy decoding
(\cref{app:bootstrap}), and byte equality is checked at both 1.7B and
4B; most secondary analyses, however, use a single draw\rev{, and the
cross-family check (Llama-3.1-8B, \cref{sec:experiments-llama}) covers
one model at the $8$B scale at which \citet{jin2026agentprimitives}
report substantially larger RoPE-related effects; consistent with their
report, the ordering gap there is the largest we measure, and larger
models may move it further}. The benchmark family is
primarily synthetic; \rev{natural-language evaluation covers
$200$ HotpotQA bridge-$k=2$ problems and $100$ MuSiQue 2-hop problems
(\cref{sec:experiments-hotpotqa}), and broader multi-document settings
(more than two supporting documents, distractor-heavy retrieval) remain
future work.}

The convergence guarantees of \LatentFragmentSet{}
(\cref{sec:method-state}) are \emph{syntactic}: the set absorbs
byte-identical re-deliveries but not semantic near-duplicates; a
semantically convergent join that also merges near-duplicate but
non-identical fragments is future work. Render is also lazy, each
newly absorbed fragment forces an $\mathcal{O}(m)$ re-render of the
$m$-fragment set at the next decode, with no incremental reuse across rounds.
This is negligible for batch inference and is the price of exact semantics
under streaming re-delivery. (The \emph{rendered} \CanonicalMerge{} operator,
taken alone, is not idempotent; the algebraic properties hold on the state,
not on the operator, \cref{sec:method-state}.)

Our byte-identity claims are verified within a single run on a fixed model
and device. The cross-party federated reading
(\cref{sec:related-federated}) carries one further condition we make
explicit, following \citet{gillespie2026crdtmerge}: bit-for-bit equality
assumes \emph{matched arithmetic}, identical model weights and a
deterministic, IEEE-754-consistent kernel path across parties (same
hardware/ISA, no nondeterministic reductions). Under heterogeneous hardware,
independently encoded fragments may differ in their low-order bits, which
would defeat content-hash absorption and exact byte-equality across replicas.
The content-determined \emph{ordering} and the set-union \emph{convergence}
algebra are unaffected, they depend on fragment identity, not on
cross-device bit-reproducibility, but byte-equality across heterogeneous
accelerators is not claimed; a tolerance-based fragment identity is the
natural remedy and is left to future work.

\rev{Our experiments are also RoPE-only. The positional encoding is fixed
by a pretrained model's weights and the method is training-free on top of
existing models, which at the scales studied are uniformly RoPE-based;
prior latent-exchange work defines its merge in RoPE terms for the same
reason. The construction itself does not depend on RoPE:
\LatentFragmentSet{} stores fragments \emph{unshifted}, so the state
layer is independent of the positional encoding and all PE handling is in
the render. For RoPE
the render applies the deterministic re-rotation of
\cref{eq:bagmerge-k}; for a model with additive attention biases in the
ALiBi style~\citep{press2022alibi}, the render would be simpler still, a
pure block concatenation, since position dependence enters at attention
time rather than in the cached K. Evaluating the merge on a
non-RoPE pretrained model is interesting future work if suitable open
models at these scales become available.}

Finally, the latent construction's primary advantage,
commutativity and byte-identity, is a structural property not
measurable in text space. \rev{The matched-budget text-fusion baseline
(\cref{sec:experiments-packllm}) quantifies the accuracy side of that
trade: shipping the fragments as text reaches the benchmark ceiling, at
the cost of violating the federated constraint; richer text-side
systems (RAG, FiD) remain uncompared.}

\section{Beyond \texorpdfstring{$N = 2$}{N = 2}}
\label{sec:beyond-n2}

\rev{Set-union transport generalizes to any arity by construction;
model-mediated composition does not automatically follow. That is the
core message of this section, established at $k = 3$: \emph{CRDT cache
merge is a transport and colocation operator, not a composition
operator}. The paper's accuracy claims remain $N = 2$; the $k = 3$
boundary delimits them and frames the open problem.}

\paragraph{The state extends exactly; the open question is composition.}
The replicated state is $N$-general by construction: set union
(\cref{eq:setunion}) is defined for any number of fragments, and the render
inherits the byte-identity we verify across all $N!$ input permutations for
$N \in \{2, 3, 4, 5\}$, $152$ permutation pairs, on synthetic tensors
(\cref{app:byte-crdt-tests}). What neither the set nor its render can
guarantee is that the \emph{model} composes the fragments they colocate. At
$k = 3$ that question has a sharp answer.

\paragraph{Transport is not composition.}
We built a $k = 3$ partitioned benchmark (three jointly-necessary fragments,
the same five families). It is sound: the full-text ceiling is
$C_0 \approx 1.00$ ($0.97$--$1.00$ across families) while every
$(k\!-\!1)$-subset baseline scores $\leq 5\%$, so any latent gap is a
\emph{composition} failure, not task difficulty. \rev{\Cref{fig:topology}
depicts four ways to combine the three latent traces;
\cref{tab:topology} in \cref{app:topology} quantifies them on the two
families that mark the poles of the effect, with full five-family
results in \cref{tab:topology-full}.} A flat $3$-way \CanonicalMerge{} (\textsc{par3})
collapses to chance on chain-structured problems, and so does a pairwise merge
\emph{tree} (\textsc{treemerge}). For the tested families, then, cache-only
colocation is insufficient and a pairwise cache-merge tree does not by itself
create the missing composition, it is not the merge arity or the
flat-versus-tree shape. Composition appears only when the model reprocesses
one or more seams: running the second thinker on the first's latent state
(\textsc{mix}), or chaining all three (\textsc{seq3}), recovers quality up to
the full-text ceiling. The merge moves latent traces into a shared cache,
deterministically and order-invariantly; composition, on this evidence,
requires the model to reprocess them. (This does not contradict Agent
Primitives' working $N = 3$ merge, which aggregates among complete voting
candidates rather than composing jointly-necessary fragments;
\cref{sec:related-bagmerge}.) This boundary is consistent with
independent recent findings that continuous-thought latent traces transport
and scaffold but do not by themselves perform the
reasoning~\citep{latenttokens2026,superposition2025}, that learned
KV-consolidation at reasoning-step boundaries outperforms plain cache
accumulation~\citep{bottlenecked2025}, and that training-free parallel-cache
synthesis underperforms until a learned calibration step is
added~\citep{liu2026parallelsynthesis}, a pattern consistent with model
reprocessing, rather than colocation, being what supplies composition.

\begin{figure}[t]
\centering
\begin{tikzpicture}[
  thinker/.style={draw, rounded corners=2pt, minimum size=5mm, inner sep=1.5pt, font=\footnotesize},
  mergeop/.style={draw, circle, inner sep=0pt, minimum size=4.5mm},
  judger/.style={draw, fill=black!8, minimum size=5mm, inner sep=2pt, font=\footnotesize\bfseries},
  medge/.style={-{Latex[length=1.4mm]}, semithick},
  sedge/.style={-{Latex[length=1.4mm]}, dashed, semithick, blue!65!black},
  plabel/.style={font=\footnotesize},
  acc/.style={font=\scriptsize},
]
\begin{scope}[xshift=0cm]
  \node[plabel] at (0.9,3.15) {(a) \textsc{par3}};
  \node[thinker] (a1) at (0.1,2.5) {$A_1$};
  \node[thinker] (a2) at (0.9,2.5) {$A_2$};
  \node[thinker] (a3) at (1.7,2.5) {$A_3$};
  \node[mergeop] (am) at (0.9,1.25) {$\uplus$};
  \node[judger]  (aj) at (0.9,0) {J};
  \draw[medge] (a1)--(am); \draw[medge] (a2)--(am); \draw[medge] (a3)--(am);
  \draw[medge] (am)--(aj);
  \node[acc] at (0.9,-0.55) {flat merge};
\end{scope}
\begin{scope}[xshift=3cm]
  \node[plabel] at (0.9,3.15) {(b) \textsc{treemerge}};
  \node[thinker] (b1) at (0.1,2.5) {$A_1$};
  \node[thinker] (b2) at (0.9,2.5) {$A_2$};
  \node[thinker] (b3) at (1.7,2.5) {$A_3$};
  \node[mergeop] (bm1) at (0.5,1.6) {$\uplus$};
  \node[mergeop] (bm2) at (1.1,0.8) {$\uplus$};
  \node[judger]  (bj) at (1.1,0) {J};
  \draw[medge] (b1)--(bm1); \draw[medge] (b2)--(bm1);
  \draw[medge] (bm1)--(bm2); \draw[medge] (b3)--(bm2);
  \draw[medge] (bm2)--(bj);
  \node[acc] at (0.9,-0.55) {pairwise tree};
\end{scope}
\begin{scope}[xshift=6cm]
  \node[plabel] at (0.95,3.15) {(c) \textsc{mix}};
  \node[thinker] (c1) at (0.25,2.5) {$A_1$};
  \node[thinker] (c2) at (0.25,1.55) {$A_2$};
  \node[thinker] (c3) at (1.7,2.5) {$A_3$};
  \node[mergeop] (cm) at (1.0,0.85) {$\uplus$};
  \node[judger]  (cj) at (1.0,0) {J};
  \draw[sedge] (c1)--(c2);
  \draw[medge] (c2)--(cm); \draw[medge] (c3)--(cm);
  \draw[medge] (cm)--(cj);
  \node[acc] at (0.95,-0.55) {one seam};
\end{scope}
\begin{scope}[xshift=9cm]
  \node[plabel] at (0.6,3.15) {(d) \textsc{seq3}};
  \node[thinker] (d1) at (0.6,2.5) {$A_1$};
  \node[thinker] (d2) at (0.6,1.66) {$A_2$};
  \node[thinker] (d3) at (0.6,0.83) {$A_3$};
  \node[judger]  (dj) at (0.6,0) {J};
  \draw[sedge] (d1)--(d2); \draw[sedge] (d2)--(d3);
  \draw[medge] (d3)--(dj);
  \node[acc] at (0.6,-0.55) {full chain};
\end{scope}
\draw[medge] (2.4,-1.45)--(2.85,-1.45);
\node[anchor=west, font=\scriptsize] at (2.9,-1.45) {render merge ($\uplus$, cache-only; non-idempotent)};
\draw[sedge] (2.4,-1.9)--(2.85,-1.9);
\node[anchor=west, font=\scriptsize] at (2.9,-1.9) {reprocessing seam (a thinker runs on another's cache)};
\draw[-{Latex[length=2mm]}, gray!70!black, thick] (0.1,-2.5)--(10.6,-2.5);
\node[font=\scriptsize] at (5.35,-2.78) {more model reprocessing (left to right)};
\end{tikzpicture}
\caption{Four ways to combine three latent traces at $k = 3$, a topology
\emph{schematic}, not an algebraic identity (accuracies in
\cref{tab:topology}). Solid nodes ($\uplus$) are cache-only \emph{render
merges} (non-idempotent; \emph{not} the idempotent set-union state join
$\sqcup$ of \cref{sec:method-state}, which we reserve for \LatentFragmentSet{}).
Dashed edges are model-mediated \emph{reprocessing seams}, where one thinker
runs with another's KV cache as a prefix. The cache-only flat (a) and tree (b)
variants both fail and collapse to the same accuracy (the
$\textsc{treemerge}\approx\textsc{par3}$ result), while the variants that add
reprocessing ((c),~(d)) recover on chain-structured tasks. The topologies
differ in more than edge type, they also change which thinker reprocesses
which prefix and the amount of model computation, so we read the figure as
locating \emph{where} reprocessing enters, not as a one-variable causal claim.}
\label{fig:topology}
\end{figure}

\paragraph{A spectrum, and a composition-limited pole.}
The recovery is a spectrum, not a switch. Pooling per-problem paired
differences across the three draws, the single-seam recovery (\textsc{mix}
over \textsc{par3}) is $+0.51$ ($95\%$ bootstrap CI $[+0.44, +0.57]$, $n =
300$) on the chain families but only $+0.14$ ($[+0.10, +0.18]$, $n = 450$) on
the symmetric ones, with a composition-limited pole, the symmetric
\textsc{constraint} family, where even a full sequential chain reaches only
$0.17 \pm 0.06$ and reprocessing barely helps. The boundary is not
specific to caches: PackLLM, our output-level fusion baseline
(\cref{sec:experiments-packllm}), also collapses at $k = 3$ ($0.04
\rightarrow 0.00$ at the longer budget), and it is not an artifact of model
capacity or seam geometry, Qwen3-4B does not recover it, and inserting
RoPE-phase gaps or boundary-marker tokens between thinker blocks only lowers
accuracy. In every case the latent traces are present; what is missing is
composition.

\paragraph{The open problem.}
This delimits the present contribution and frames the next. At $N = 2$, a
content-determined render over a set-union state gives order-invariant,
convergent latent merging at no measurable quality cost. At $k > 2$, the open
problem is to recover the integration that, here, only model reprocessing
provides, while preserving as much of that order-invariant state
discipline as possible. We pursue that direction separately.

\section{Conclusion}
\label{sec:conclusion}

We began from a concrete flaw in how multi-agent latent reasoning
composes KV-cache state: the concat-with-RoPE operator implicit in prior
work, \BagMerge{}, is non-commutative, and the best input ordering is
unpredictable across regime, latent budget, model scale\rev{, and model
family}.
\CanonicalMerge{} removes the decision by fixing the layout on content
rather than caller order, rendering the merged cache byte-identical under
any permutation of the inputs, a property we verify algorithmically up to
five agents and bit-for-bit on the real KV state of Qwen3-1.7B and 4B.
Separating the replicated state from its decode-time layout then lifts
this from a property of an operator to a property of a durable object: a
set of content-addressed latent fragments whose set-union merge is a
state-based CvRDT, of which \CanonicalMerge{} is the deterministic
render. On a benchmark built so that the merge sits on the critical path,
\CanonicalMerge{} matches the best fixed ordering in each regime without
needing to select it (multi-seed, paired-bootstrap, Holm-corrected),
operates in a regime distinct from output-level fusion (PackLLM trails by
$45$ points at matched budget), and, under repeated fragment
re-delivery, holds accuracy and cache size flat where a provenance-free
operator degrades and grows linearly with repeated delivery.

Two implications follow. First, treating inter-agent cache exchange as a
convergent replicated state gives federated or query-blind latent
inference structural rather than coordinative guarantees: independent
parties can contribute pre-encoded state in any order, under any delivery
schedule, and with duplicates, and obtain the same rendered context,
without a coordinator dictating order or deduplicating messages. This
places cross-agent KV-cache merging as an inference-time primitive on
equal footing with the within-sequence KV-cache compression the systems
literature already treats as first-class. Second, the approach has a
clear boundary: beyond two agents, cache merging transports and colocates
latent traces but does not by itself compose them; a pairwise merge tree
buys nothing over a flat merge, and in the tested families can even lose;
composition is recovered only when the model reprocesses the colocated
state through attention. Turning transport into composition, without
paying the full cost of sequential reprocessing, is the problem we leave
to future work.

\rev{\section*{Broader Impact Statement}

Cache-level exchange in federated or query-blind settings has implications
beyond convergence, and none of the structural guarantees in this paper
mitigate them. First, \emph{privacy}: KV caches can encode sensitive
content of the source fragments, and latent exchange must not be presented
as privacy-preserving by default; cross-party cache exchange carries
confidentiality and inversion risk, and neither convergence nor content
hashing reduces it. Second, \emph{authenticity and poisoning}: content
addressing provides identity, not authenticity or safety. A malicious
contributor can submit poisoned fragments, and because the position-$0$
prefix slot is privileged (\cref{sec:method-canonical}), an adversary
could craft fragments with inflated K-norm to capture it; any
deterministic ordering admits an analogous, if harder to target, variant.
Provenance, signatures, and access control are orthogonal mechanisms that
a deployment must add. Third, \emph{linkability and retention}: fragment
hashes are stable identifiers, so retaining them enables cross-request
linkage of contributions; deployments must weigh this against the
deduplication benefit and set retention policy accordingly. Finally,
cache-level merging changes \emph{where} injected content enters a
pipeline: it arrives as encoded state rather than as readable text, which
can obscure provenance relative to text-level retrieval and deserves
scrutiny when fragments come from independent parties.}

\bibliography{references}
\bibliographystyle{tmlr}

\appendix

\section{Synthetic-tensor byte-CRDT tests}
\label{app:byte-crdt-tests}

We first establish the byte-CRDT property of the render on synthetic
state, with no model inference, so that permutation invariance is
isolated from any modelling noise. The test
(\texttt{tests/test\_canonical\_n.py}) builds legacy-style KV caches of
small random Gaussian tensors, choosing the per-thinker scales so that
the mean K-norms at the routing layer are cleanly separable and the
canonical order is therefore unambiguous. Equality is checked at the bit
level (\texttt{torch.equal} on the raw tensors, not approximate
\texttt{allclose}). Three properties are verified.

\paragraph{Permutation invariance at \texorpdfstring{$N > 2$}{N > 2}.}
For $N \in \{2, 3, 4, 5\}$ we render every one of the $N!$ input
orderings and assert each is byte-identical to a fixed reference
ordering, $2 + 6 + 24 + 120 = 152$ permutations in total, all
identical. This extends the commutativity of
\cref{eq:canonical-commute}, proved in the main text for $N = 2$, to
higher arity: the content-determined layout selects the same slot for
each fragment regardless of the order in which the caches are presented.

\paragraph{Reduction to the \texorpdfstring{$N = 2$}{N = 2} operator.}
Called with two caches, the $N$-fold path
\texttt{novelty\_concat\_sym\_layout\_n} produces output byte-identical
to the dedicated two-input strategy
\texttt{novelty\_concat\_sym\_layout}, confirming that the batch
generalisation introduces no regression at the arity used for every
headline number.

\paragraph{Canonical-ordered concatenation (the headline configuration).}
In the configuration used throughout the paper the merge is a pure
canonical-ordered concatenation: the rendered length equals the sum of
the input lengths ($12, 18, 24$ for $N = 2, 3, 4$ at six positions per
thinker), with no position dropped, and a further $32$ permutations
($2! + 3! + 4!$) are all byte-identical. \rev{With the headline threshold
$\tau = 2.0$ this is provable: positions are suppressed only when
inter-thinker cosine agreement exceeds $\tau$, and cosine similarity is
bounded by $1$, so the test never fires and every position of every
fragment is kept (the dedicated test
\texttt{test\_high\_threshold\_no\_dedup} asserts exactly this). Byte-equality
across input orders is exact because the merge performs no
\emph{cross-fragment} arithmetic: each output element is a deterministic
function of exactly one input fragment and its content-determined canonical
rank. The prefix block keeps its original rotation and every subsequent
block's K is re-rotated by its cumulative slot offset
(\cref{eq:bagmerge-k}), a deterministic elementwise kernel; since the rank
does not depend on input order, both orders execute the identical sequence
of tensor operations and no floating-point non-associativity is introduced
by reordering} (\cref{app:realmodel-byte-equality}; the IEEE-754 condition
across replicas is noted in \cref{sec:limitations}).

Together these establish the render's input-order independence and
deterministic layout on controlled state, separately from the model;
\cref{app:realmodel-byte-equality} repeats the permutation check on real
Qwen3 caches, and the state-level formalisation that makes this a CvRDT
join rather than a property of the operator alone is given in
\cref{sec:method-state}.

\section{Bootstrap CIs and McNemar paired tests}
\label{app:bootstrap}

This appendix backs the accuracy comparisons of
\cref{sec:experiments-stats} with a multi-seed analysis. We re-ran the
$N = 2$ matrix, \CanonicalMerge{} (CM), \BagMerge{} in its default and
swapped orderings (BG, BGSW), and the PackLLM output-fusion baseline, in
both regimes under greedy decoding, across three independent problem draws
(data seeds $\{42, 7, 100\}$, $n = 100$ each, at the $40$-step budget and the
$2048$-token headline generation cap). Greedy decoding is deterministic, so the
relevant variance is the problem draw: we report per-cell accuracy as the mean
$\pm$ s.d.\ across draws with a pooled $n = 300$ Wilson interval
(\cref{tab:nstats-cells}), and every method comparison as a per-problem paired
bootstrap ($10\,000$ resamples) pooled across draws, with an exact McNemar $p$
and Holm correction over the seven-contrast family
(\cref{tab:nstats-contrasts}). Analysis script:
\texttt{scripts/analyze\_nstats.py}. The single-seed $20$/$40$-step bootstrap
that \cref{tab:mcnemar} summarises is retained at
\texttt{runs/bootstrap\_results\{,\_20step\}.md}.

\begin{table}[t]
\centering
\small
\begin{tabular}{llccc}
\toprule
method & regime & per-seed ($42/7/100$) & mean $\pm$ s.d. & pooled $95\%$ CI \\
\midrule
\CanonicalMerge{}   & known & $0.96 / 0.95 / 0.93$ & $0.95 \pm 0.02$ & $[0.92, 0.97]$ \\
\CanonicalMerge{}   & blind & $0.86 / 0.93 / 0.86$ & $0.88 \pm 0.04$ & $[0.84, 0.91]$ \\
\BagMerge{} default & known & $0.88 / 0.89 / 0.88$ & $0.88 \pm 0.01$ & $[0.84, 0.91]$ \\
\BagMerge{} default & blind & $0.89 / 0.88 / 0.89$ & $0.89 \pm 0.01$ & $[0.85, 0.92]$ \\
\BagMerge{} swap    & known & $0.96 / 0.97 / 0.93$ & $0.95 \pm 0.02$ & $[0.92, 0.97]$ \\
\BagMerge{} swap    & blind & $0.81 / 0.83 / 0.82$ & $0.82 \pm 0.01$ & $[0.77, 0.86]$ \\
PackLLM             & known & $0.51 / 0.49 / 0.50$ & $0.50 \pm 0.01$ & $[0.44, 0.56]$ \\
\bottomrule
\end{tabular}
\caption{Per-cell accuracy across three problem draws (Qwen3-1.7B, $k = 2$,
greedy, $n = 100$ per draw; pooled $95\%$ Wilson CI at $n = 300$). Per-seed
values are ordered by data seed $42 / 7 / 100$. \CanonicalMerge{} and the
best-of-two \BagMerge{} ordering coincide in each regime (swap in query-known,
default in query-blind), and the low across-draw s.d.\ shows the numbers are
draw-stable except for query-blind \CanonicalMerge{}, whose single-draw swing
motivates the paired analysis below.}
\label{tab:nstats-cells}
\end{table}

\begin{table}[t]
\centering
\small
\begin{tabular}{lccc}
\toprule
contrast & $\Delta$\,acc [$95\%$ CI] & McNemar $p$ & Holm \\
\midrule
CM ${}-{}$ BG (known)      & $+0.063$ [$+0.037, +0.093$] & $<0.001$ & reject \\
CM ${}-{}$ BGSW (known)    & $-0.007$ [$-0.040, +0.027$] & $0.839$  & n.s. \\
CM ${}-{}$ BG (blind)      & $-0.003$ [$-0.043, +0.037$] & $1.000$  & n.s. \\
CM ${}-{}$ BGSW (blind)    & $+0.063$ [$+0.037, +0.093$] & $<0.001$ & reject \\
BG ${}-{}$ BGSW (known)    & $-0.070$ [$-0.113, -0.027$] & $0.003$  & reject \\
BG ${}-{}$ BGSW (blind)    & $+0.067$ [$+0.020, +0.113$] & $0.011$  & reject \\
CM ${}-{}$ PackLLM (known) & $+0.447$ [$+0.387, +0.503$] & $<0.001$ & reject \\
\bottomrule
\end{tabular}
\caption{Per-problem paired bootstrap ($10\,000$ resamples, $n = 300$ pooled
across the three draws), exact McNemar $p$, and Holm-corrected decision over the
seven-contrast family. \CanonicalMerge{} ties the best ordering in each regime
(CM${}-{}$BGSW query-known and CM${}-{}$BG query-blind both straddle zero) and
beats the worse ordering ($+0.063$, Holm-significant); \BagMerge{}'s
order-sensitivity (BG${}-{}$BGSW) is significant and reverses sign between
regimes.}
\label{tab:nstats-contrasts}
\end{table}

Two readings follow. First, \CanonicalMerge{} \emph{ties the best fixed
ordering in each regime}, CM${}-{}$BGSW (query-known) and CM${}-{}$BG
(query-blind) both straddle zero and are non-significant after Holm, while it
\emph{beats the worse ordering} in both regimes by $+0.063$
($p < 0.001$, Holm-significant). This is the ``matches the best ordering without
selecting it'' claim, now with the between-draw variance folded in. Second,
\BagMerge{}'s order-sensitivity (BG${}-{}$BGSW) is significant in both regimes
and \emph{reverses sign} between them ($-0.070$ query-known, $+0.067$
query-blind), so no single ordering is safe across regimes, precisely the
choice \CanonicalMerge{} removes. The output-fusion baseline PackLLM sits
$+0.447$ [$+0.387, +0.503$] below \CanonicalMerge{} even at the matched
$2048$-token budget, a gap an order of magnitude larger than any ordering effect
(\cref{app:packllm-details}).

\rev{\paragraph{Family-stratified accuracies.} \Cref{tab:family-strat}
breaks the pooled matrix down by benchmark family (three seeds pooled,
$n = 60$ per family per cell). Two families are saturated for every
method (\textsc{lookup}, \textsc{recipe}, and \textsc{discount} nearly
so); the ordering effects concentrate in \textsc{constraint} and
\textsc{neonym}, and the best fixed ordering flips between them.}

\rev{\begin{table}[t]
\centering
\small
\begin{tabular}{ll ccccc c}
\toprule
Method & Regime & \textsc{constr.} & \textsc{disc.} & \textsc{lookup} & \textsc{neonym} & \textsc{recipe} & overall \\
\midrule
CM   & known & $0.87$ & $1.00$ & $1.00$ & $0.87$ & $1.00$ & $0.947$ \\
BG   & known & $0.55$ & $1.00$ & $1.00$ & $0.87$ & $1.00$ & $0.883$ \\
BGSW & known & $0.97$ & $1.00$ & $1.00$ & $0.80$ & $1.00$ & $0.953$ \\
\midrule
CM   & blind & $0.60$ & $1.00$ & $1.00$ & $0.82$ & $1.00$ & $0.883$ \\
BG   & blind & $0.62$ & $1.00$ & $1.00$ & $0.82$ & $1.00$ & $0.887$ \\
BGSW & blind & $0.60$ & $0.97$ & $1.00$ & $0.53$ & $1.00$ & $0.820$ \\
\bottomrule
\end{tabular}
\caption{Family-stratified accuracy of the $N = 2$ matrix (seeds
$42/7/100$ pooled, $n = 60$ per family per cell). The ordering effects
live in \textsc{constraint} and \textsc{neonym}; the best fixed
\BagMerge{} ordering is family-dependent (swap wins \textsc{constraint}
query-known but loses \textsc{neonym}), and in query-blind the swap
ordering collapses on \textsc{neonym} ($0.53$) while \CanonicalMerge{}
holds $0.82$.}
\label{tab:family-strat}
\end{table}}

\rev{\paragraph{Development/evaluation separation.} For Qwen3-1.7B, the
routing layer $\ell^\star = 15$ was chosen during method development,
before the headline matrix was run; the layer sweep of
\cref{sec:experiments-layer} shares the seed-$42$ draw with the headline
table and is therefore reported as sensitivity, not selection, and the
finer sweep on the held-out seed-$7$ draw validates the choice on
independent data ($\ell^\star$ was fixed before that sweep). Seed $7$ is
also one of the three evaluation draws of this appendix; the held-out
sweep selected nothing, but we note the reuse. For Qwen3-4B,
$\ell^\star = 24$ was selected by a coarse sweep over
$\{15, 19, 24\}$ on the first $50$ problems of the default seed-$42$
draw, the same draw evaluated in \cref{tab:headline-4b}; selection and
evaluation data therefore overlap in that table. For this reason all six
4B cells are also evaluated on the disjoint seed-$7$ draw with
$\ell^\star = 24$ frozen, reported in \cref{sec:experiments-scale}.}

\section{Real-model byte-equality on the headline configuration}
\label{app:realmodel-byte-equality}

This appendix is the real-model companion to the synthetic byte-CRDT
test of \cref{app:byte-crdt-tests}: it verifies that
\cref{eq:canonical-commute} holds not only on controlled random tensors
but on KV caches produced by an actual model forward pass, in the exact
headline configuration. The test
(\texttt{tests/test\_byte\_equality\_real\_model.py}) runs Qwen3-1.7B on
the first partitioned $k = 2$ problem in the query-blind regime, under
the headline merge settings (\texttt{novelty\_concat\_sym\_layout},
$\tau = 2.0$, $\ell^\star = 15$, $40$ latent steps). It encodes the two
fragments under both orderings, order~A
$(\text{thinker}_0 \leftarrow t_a,\ \text{thinker}_1 \leftarrow t_b)$ and
the swapped order~B
$(\text{thinker}_0 \leftarrow t_b,\ \text{thinker}_1 \leftarrow t_a)$,
the \texttt{-{}-swap\_partitioned\_thinkers} setting, runs each
thinker's forward pass plus latent steps to produce its cache, and
merges each ordering into \texttt{merged\_A} and \texttt{merged\_B}.
\rev{The swap is a \emph{pure operand permutation}, not a change of
cache contents, for two reasons. First, for partitioned split-view tasks
the implementation overrides any thinker-role diversity setting and
assigns every thinker the identical \texttt{planner} role; the thinker
prompt is built from the role and the fragment text only, with no
thinker index entering it, so under the swap thinker~$0$ encodes exactly
the prompt thinker~$1$ encodes in the default order. Second, we verify
this end to end with an \emph{encode-once} variant
(\texttt{tests/test\_encode\_once\_permutation.py}): each fragment is
encoded exactly once and only the operand list handed to the merge is
permuted. On real Qwen3-1.7B caches the swap prompts are string-identical
to the default prompts exchanged; \CanonicalMerge{} on the fixed operand
pair is byte-identical under swap on all $28$ layers (both strategy
variants and their $N=2$ reduction); and \BagMerge{} on the \emph{same
fixed bytes} differs on all $28$ layers, isolating the order sensitivity
to the operator with zero re-encoding. Results replicate on both
\textsc{mps} and \textsc{cuda} backends, and at the second model scale
(Qwen3-4B, $\ell^\star = 24$: byte-identical on $36/36$ layers for
\CanonicalMerge{}, all $36$ differing for \BagMerge{}).}

Equality is checked at the byte level. Each layer's K and V tensors are
moved to CPU, made contiguous, and reinterpreted as \texttt{uint8} over
the raw \texttt{bfloat16} storage, numpy has no native
\texttt{bfloat16}, and we want a bit-pattern comparison rather than a
dtype-aware one, then SHA-1 hashed. All $28$ layers of K and all $28$
layers of V hash byte-identical between \texttt{merged\_A} and
\texttt{merged\_B}, establishing \cref{eq:canonical-commute} on real
model state for the headline benchmark configuration. This is the
single strongest piece of evidence for commutativity at $N = 2$: the
rendered cache is bit-for-bit independent of input order, so the
downstream decode, and therefore every accuracy number in
\cref{sec:experiments}, cannot depend on which fragment was presented
first. The check is exact (\texttt{torch.equal} / SHA-1, not
\texttt{allclose})\rev{; exact equality is attainable because the merge
performs no cross-fragment arithmetic and applies to each block only a
rotation determined by its content-derived canonical rank
(\cref{app:byte-crdt-tests}), so reordering the inputs changes no tensor
operation. Reproducing the rotation arithmetic bit-for-bit \emph{across
replicas} is the IEEE-754 caveat discussed in \cref{sec:limitations}.}

\rev{Because every output byte is a function of one fragment and its
content-determined rank,} byte-identity
is a property of the layout rather than of the model or its width: the
synthetic test (\cref{app:byte-crdt-tests}) confirms it across tensor
shapes and arities $N \le 5$ independently of any architecture. We run
the same real-model check at the second model scale: on Qwen3-4B
(with $\ell^\star = 24$), all $36$ layers of K and V hash byte-identical
between the two orderings, confirming \cref{eq:canonical-commute} on real
4B state as well; the accuracy-level replication is reported in
\cref{sec:experiments-scale}.

\section{PackLLM adaptation details}
\label{app:packllm-details}

\paragraph{Construction.} We adapt PackLLM~\citep{mavromatis2024packllm}
to the single-base-model, $N$-thinker setting so that it shares
\CanonicalMerge{}'s inputs and differs only in \emph{where} the
combination happens. The two methods are identical through Phase~1:
each thinker independently encodes its fragment and runs the same
latent steps, producing a cache $\mathcal{C}_i$. They then diverge.
\CanonicalMerge{} merges the caches and decodes once from the merged
prefix (\cref{sec:method}); our PackLLM keeps the $N$ caches separate
and fuses at the logits:
\begin{enumerate}
\item \emph{Phase 2 (weighting).} The judger's chat-templated prompt is
encoded on top of each thinker's cache $\mathcal{C}_i$; we record the
perplexity $\mathrm{PPL}_i$ of that prompt under each cache and the
next-token logits at its final position. Weights are
$\lambda_i = \operatorname{softmax}(-\log \mathrm{PPL}_i / \tau)$
(``PackLLMsim''); a greedy grid variant (``PackLLMopt'', $N=2$ only) is
also implemented but not reported as it does not change the conclusion.
\item \emph{Phase 3 (fused decode).} At each step the fused
distribution is $\operatorname{softmax}\!\big(\sum_i \lambda_i\,
\mathbf{z}_i\big)$ over per-thinker logits $\mathbf{z}_i$ (averaging is
pre-softmax, matching PackLLM's Eq.~2); the sampled token is appended to
\emph{all} $N$ caches so they advance in lockstep, and the per-thinker
logits are recomputed for the next step.
\end{enumerate}
We use the judger-prompt PPL as the weighting signal (the
PackLLM-faithful choice), sample at temperature $0.7$ / $\texttt{top\_p}
= 0.95$ to match \CanonicalMerge{} (PackLLM's original uses argmax), and
fuse pre-softmax. Reference implementation:
\texttt{methods/packllm\_baseline.py}, which subclasses the
\CanonicalMerge{} method to reuse Phase~1 verbatim.

\paragraph{The PackLLM-faithful setting and the one-hot limit both
fail.} The weighting is near-uniform because the two thinkers'
judger-prompt perplexities are close: at $\tau = 1.0$, $\lambda_0$ has
mean $0.49$, s.d.\ $0.04$ in the query-known arm (mean PPLs $234$ vs.\
$223$) and mean $0.52$, s.d.\ $0.07$ in query-blind. Driving
$\tau \to 0$ collapses the weighting to one-hot on the lower-PPL thinker
(query-blind, $\tau = 0.01$: $\lambda_0$ bimodal at $\{0, 1\}$), which
decodes from a single partition and lands at the partitioned-blind floor
of $0.01$. With the two endpoints failing and the perplexities so nearly
equal, it is unlikely that any intermediate $\tau$ turns them into a
useful integration signal.

\paragraph{Truncation cannot account for the gap.} A reviewer might
worry that PackLLM's low accuracy is a generation-length artifact, since
we cap decoding at $512$ tokens. We bound this. Per-problem decode
lengths and a truncation flag ($\texttt{decode\_len} \geq 512$) are
recorded for every problem
(\texttt{runs/packllm\_overnight/per\_sample\_telemetry.csv}).
\Cref{tab:packllm-trunc} gives the worst-case ceiling under the
(maximally generous) assumption that \emph{every} truncated-and-wrong
problem would become correct with unlimited budget. Even then PackLLM
tops out at $0.89$ (query-known) and $0.70$ (query-blind), still below
\CanonicalMerge{}'s $0.94 / 0.85$ at the matched $40$-step budget. The
concern is then settled directly at $k = 2$: re-running PackLLM at the
matched $2048$-token generation budget across three problem draws
($n = 300$) gives $0.50 \pm 0.01$ in query-known, still $+0.447$
[$+0.387, +0.503$] below \CanonicalMerge{} (\cref{app:bootstrap}), so
the gap is a fusion-level limitation, not a generation-length artifact.
The $k=3$ follow-up corroborates it: re-running PackLLM at
$\texttt{max\_new\_tokens} = 2048$ \emph{lowered} accuracy from $0.04$
to $0.00$ (\cref{sec:beyond-n2}), the extra budget lets the model
converge more confidently on ``insufficient information'' rather than
synthesise an answer.

\begin{table}[t]
\centering
\small
\begin{tabular}{l c c c c}
\toprule
arm & actual acc & truncated@512 (wrong) & decode len (mean) & ceiling \\
\midrule
query-known, $\tau{=}1.0$ & $0.38$ & $54$ ($51$) & $417$ & $0.89$ \\
query-blind, $\tau{=}1.0$ & $0.34$ & $44$ ($36$) & $397$ & $0.70$ \\
query-blind, $\tau{=}0.01$ & $0.01$ & $29$ ($28$) & $358$ & $0.29$ \\
\bottomrule
\end{tabular}
\caption{PackLLM truncation upper bound ($100$ problems each,
Qwen3-1.7B). ``ceiling'' assumes every truncated-and-wrong problem
flips to correct under unlimited generation budget; it remains below
\CanonicalMerge{}'s $0.94$ (query-known) and $0.85$ (query-blind).}
\label{tab:packllm-trunc}
\end{table}

\section{Same-data motivation}
\label{app:same-data-motivation}

A natural alternative to the partitioned benchmark
(\cref{sec:benchmark}) is a \emph{same-data} design: two thinkers see
the same problem and are diversified only by role prompts or injected
noise. We did not use it, because it cannot isolate the merge operator's
contribution. When every thinker receives the complete query, a single
thinker already has everything needed to solve the problem, so the merge
is not on the critical path, it is an optional add-on to an
already-sufficient input. Any accuracy difference between a two-thinker
merge and a single thinker then conflates the operator's contribution
with two confounds: extra inference compute spent on the same problem,
and a regularization-like effect from combining noisy copies of one
input. Our own same-data pilot bore this out: the best-looking merge
configuration was behaviourally indistinguishable from a single (noisy)
thinker, the second cache contributed essentially nothing to the
decoded answer, and the small apparent lift tracked the extra forward
pass rather than any integration of a second source.

The partitioned construction removes the confound by design. Each
problem is split into fragments $t_a$ and $t_b$ that are
\emph{individually insufficient} and \emph{jointly necessary}: neither
alone determines the answer, but together they fix it uniquely
(\cref{sec:benchmark-construction}). A correct answer therefore
\emph{requires} integrating both caches, putting the merge operator
squarely on the critical path. The empirical validation confirms the
separation (\cref{sec:benchmark-validation}): the single-fragment
baselines floor near chance ($C_1 = 5\%$ for $t_a$ alone, $C_2 = 2\%$
for $t_b$ alone) while the full-text ceiling is $C_0 = 100\%$, so the
task is solvable precisely when both fragments are present. The gap
between the single-fragment floor and the merged accuracy is then
attributable to the merge itself, with ``extra compute on the same
query'' eliminated as an alternative explanation by construction.

\section{Full five-family \texorpdfstring{$k = 3$}{k=3} topology results}
\label{app:topology}

\Cref{tab:topology-full} reports the $k = 3$ topology probe across all five
families; \cref{tab:topology} shows the two poles
(\textsc{neonym}, \textsc{constraint}).

\begin{table}[t]
\centering
\caption{The $k = 3$ topology probe, on the two families that mark the poles
of the effect (Qwen3-1.7B, greedy; \emph{across-draw mean $\pm$ sd} over $3$
independent problem draws, data seeds $\{42, 7, 100\}$, $n = 50$/family/draw;
full five-family results in \cref{tab:topology-full}). Both problems are
near-perfectly solvable from full text ($C_0 \approx 1.00$), so every latent
gap is composition, not difficulty. A flat $3$-way merge (\textsc{par3}) and a
pairwise merge tree (\textsc{treemerge}) both collapse; introducing model
reprocessing at one seam (\textsc{mix}) or throughout (\textsc{seq3}) recovers
it.}
\label{tab:topology}
\begin{tabular}{lcc}
\toprule
 & NEONYM (chain) & CONSTRAINT (symmetric) \\
\midrule
$C_0$ (full text) & $1.00 \pm 0.00$ & $0.97 \pm 0.03$ \\
\midrule
\textsc{par3} --- $3$-way merge $M(C_1, C_2, C_3)$ & $0.07 \pm 0.02$ & $0.03 \pm 0.01$ \\
\textsc{treemerge} --- pairwise $M(M(C_1, C_2), C_3)$ & $0.06 \pm 0.02$ & $0.01 \pm 0.01$ \\
\textsc{mix} --- one attention seam, then $2$-way merge & $0.73 \pm 0.03$ & $0.03 \pm 0.01$ \\
\textsc{seq3} --- sequential chain, no merge & $0.97 \pm 0.01$ & $0.17 \pm 0.06$ \\
\bottomrule
\end{tabular}
\end{table} All cells are across-draw mean $\pm$ sd
over three independent problem draws (data seeds $\{42, 7, 100\}$),
Qwen3-1.7B, greedy decoding, $n = 50$ per family per draw. All
cells, including $C_0$, are over the three draws; the lowest single-draw
full-text ceiling is $0.94$ (\textsc{constraint}), so every family is
near-perfectly text-solvable and the latent gaps are composition, not task
difficulty.

\begin{table}[h]
\centering
\small
\begin{tabular}{llccccc}
\toprule
family & type & $C_0$ & \textsc{par3} & \textsc{mix} & \textsc{seq3} & \textsc{treemerge} \\
\midrule
\textsc{neonym}     & chain & $1.00 \pm 0.00$ & $0.07 \pm 0.02$ & $0.73 \pm 0.03$ & $0.97 \pm 0.01$ & $0.06 \pm 0.02$ \\
\textsc{relay}      & chain & $1.00 \pm 0.00$ & $0.06 \pm 0.04$ & $0.41 \pm 0.07$ & $0.32 \pm 0.03$ & $0.06 \pm 0.04$ \\
\midrule
\textsc{constraint} & symm. & $0.97 \pm 0.03$ & $0.03 \pm 0.01$ & $0.03 \pm 0.01$ & $0.17 \pm 0.06$ & $0.01 \pm 0.01$ \\
\textsc{recipe}     & symm. & $1.00 \pm 0.00$ & $0.73 \pm 0.03$ & $0.96 \pm 0.02$ & $0.99 \pm 0.01$ & $0.71 \pm 0.02$ \\
\textsc{discount}   & symm. & $0.99 \pm 0.02$ & $0.23 \pm 0.11$ & $0.41 \pm 0.09$ & $0.49 \pm 0.06$ & $0.11 \pm 0.06$ \\
\bottomrule
\end{tabular}
\caption{Full five-family $k = 3$ topology probe (across-draw mean $\pm$ sd,
$3$ draws, greedy, $n = 50$). \textsc{treemerge} never exceeds \textsc{par3}
on any of the five families (colocation is not composition): it matches
\textsc{par3} on the chains and on \textsc{recipe}, and on \textsc{discount}
falls below it ($0.11$ vs.\ $0.23$), so the pairwise merge tree can even lose
relative to the flat merge. The two chain
families collapse under \textsc{par3} and recover under reprocessing; the
symmetric families span a composition-limited pole (\textsc{constraint}, where
no topology helps), through \textsc{recipe} (whose $3$-way merge already mostly
works), to a gradient (\textsc{discount}). \textsc{relay}'s
$\textsc{seq3} < \textsc{mix}$ is an ordering artifact: its default chain
places the non-commutative pipeline's input last.}
\label{tab:topology-full}
\end{table}

\section{Reproducibility commands}
\label{app:repro}

The code artifact is available at
\url{https://anonymous.4open.science/r/cache-merge-artifact-2F54}; it
contains the exact code state that produced every number in this paper,
with a \texttt{REPRODUCING.md} mapping each section and table to its
driver command. Per-run logs are written to
\texttt{runs/}, and the final metrics are the JSON object on the last
line of each log. The headline conditions reproduce with:

\small
\begin{verbatim}
# C0 ceiling (single agent, both fragments visible)
python run.py --method baseline --task partitioned \
  --partitioned_view full --model_name Qwen/Qwen3-1.7B \
  --max_samples 100 --device mps

# BagMerge (prior-work baseline): concat + RoPE re-encode
python run.py --method latent_mas_crdt --num_thinkers 2 \
  --task partitioned --partitioned_view split \
  --merge_strategy concat_rerope \
  --model_name Qwen/Qwen3-1.7B --max_samples 100 \
  --max_new_tokens 1024 --latent_steps 40 --prompt sequential \
  --device mps --device2 mps --generate_bs 1

# CanonicalMerge (ours), k = 2
python run.py --method latent_mas_crdt --num_thinkers 2 \
  --task partitioned --partitioned_view split \
  --merge_strategy novelty_concat_sym_layout \
  --novelty_threshold 2.0 --novelty_layer 15 \
  --model_name Qwen/Qwen3-1.7B \
  --max_samples 100 --max_new_tokens 1024 --latent_steps 40 \
  --prompt sequential --device mps --device2 mps --generate_bs 1

# CanonicalMerge (ours), k = 3 (partitioned_k3 dataset)
python run.py --method latent_mas_crdt --num_thinkers 3 \
  --task partitioned_k3 --partitioned_view split \
  --merge_strategy novelty_concat_sym_layout_n \
  --novelty_threshold 2.0 --novelty_layer 15 \
  --model_name Qwen/Qwen3-1.7B \
  --max_samples 100 --max_new_tokens 1024 --latent_steps 40 \
  --prompt sequential --device mps --device2 mps --generate_bs 1
\end{verbatim}
\normalsize

\rev{The commands set no \texttt{-{}-diversity\_mode}: the flag is a
no-op for partitioned split-view tasks, where the implementation
overrides it and assigns every thinker the identical \texttt{planner}
role (thinker diversity comes solely from the input fragments; see
\cref{app:realmodel-byte-equality}).}
The query-blind / federated arm adds \texttt{-{}-query\_judger\_only};
the order-sensitivity foil for \BagMerge{} adds
\texttt{-{}-swap\_partitioned\_thinkers}; the second model scale sets
\texttt{-{}-model\_name Qwen/Qwen3-4B} with
\texttt{-{}-novelty\_layer 24}; greedy decoding is
\texttt{-{}-temperature 0}.

\paragraph{Versions.} Qwen3-1.7B (and Qwen3-4B) in \texttt{bfloat16} on
Apple MPS; MacBook Pro M3, $24$\,GB unified memory; PyTorch $2.11.0$,
Transformers $4.55.4$. Seed $42$ throughout unless a data seed is
stated.

\paragraph{Note on the legacy CLI flag.} The implementation still exposes
the historical strategy name \texttt{novelty\_concat\_sym\_layout(\_n)} and
a \texttt{-{}-novelty\_threshold} argument. All reported runs set this
legacy threshold to $2.0$, which makes the strategy exactly the
canonical-ordered concatenation of \cref{sec:method-canonical}; the
threshold is not part of the method and is disclosed only so the CLI
invocations above are self-explanatory.

\end{document}